\documentclass[aps,prd,nofootinbib,showkeys,floatfix,twocolumn]{revtex4-1}
\usepackage{amsmath}
\usepackage{amssymb,ulem}
\usepackage{graphicx}
\usepackage{xspace}
\usepackage{bbm} 
\usepackage{color}
\usepackage{colortbl}
\usepackage[utf8]{inputenc}
\usepackage{cancel}
\usepackage{slashed}
\usepackage{soul}
\usepackage{xcolor,colortbl}

\definecolor{mygray}{gray}{0.8}

\definecolor{mkgreen}{rgb}{0.2,.70,.3}
\definecolor{myblue}{cmyk}{0.65, 0.37, 0.0, 0.19}

\newcommand{\lam}{\lambda}

\usepackage[pdftex,colorlinks,linkcolor=blue,citecolor=blue,urlcolor=blue]{hyperref} 

\definecolor{tobycolour}{rgb}{.6,.0,.4}

\begin{document}
\vspace{1cm}

\title{\Large Nucleon decay in the R-parity violating MSSM }

\hfill \parbox{5cm}{\vspace{ -1cm } \flushright BONN-TH-2020-13}

\newcommand{\AddrBonn}{%
Bethe Center for Theoretical Physics \& Physikalisches Institut der 
Universit\"at Bonn,\\ Nu{\ss}allee 12, 53115 Bonn, Germany}
\newcommand{\AddrDESY}{%
	DESY, Notkestra{\ss}e  85,  22607  Hamburg, Germany}
\newcommand{\AddrDamascus}{%
      Physics Department, HIAST, P.O. Box 31983, Damascus, Syria}

\author{Nidal Chamoun} \email{nidal.chamoun@hiast.edu.sy}
\affiliation{\AddrDamascus}

\author{Florian Domingo} \email{domingo@th.physik.uni-bonn.de}
\affiliation{\AddrBonn}

\author{Herbert K. Dreiner} \email{dreiner@uni-bonn.de}
\affiliation{\AddrBonn}


\begin{abstract}
We present a reanalysis of nucleon decay in the context of the R-parity violating MSSM, updating bounds
  on R-parity violating parameters against recent experimental and lattice results. We pay  particular
  attention to the derivation of these constraints and specifically to the hadronic matrix elements,
  which usually stand as the limiting factor in order to derive reliable bounds, except for these few
  channels that have been studied on the lattice.
\end{abstract}

\maketitle

\section{Introduction}
The question of matter stability emerged sixty years ago from the realization that the observed baryon
asymmetry of the Universe \cite{Canetti:2012zc} required a violation of those symmetries forbidding
proton decay \cite{Sakharov:1967dj}. While baryon number $B$ is accidentally conserved in the
Standard Model (SM) at the perturbative level (as well as lepton number $L$), it is an anomalous
symmetry and thus broken by effects such as instanton or sphaleron processes
\cite{tHooft:1976snw,Manton:1983nd,Dreiner:1992vm}. On the other hand, $B$ (or $L$) violation could
reach more dramatic proportions in constructions of new physics such as Grand-Unified Theories (GUT)
\cite{Georgi:1974sy} or supersymmetric (SUSY) extensions of the SM \cite{Nilles:1983ge,Haber:1984rc},
where the Lagrangian density is not even classically $B$- (or \mbox{$L$-)invariant}. In the former case, the
typical pattern of proton decay is imprinted in its mediation by the gauge bosons of the extended gauge
group \cite{Langacker:1980js,Raby:2002wc,Nath:2006ut,Ellis:2019fwf}. In the second case, $B$- or
$L$-conservation is conditioned to the explicit enforcement of these symmetries as a model-building
ingredient. 

The usual assumption in the Minimal Supersymmetric Standard Model (MSSM) consists in
applying an $R$-parity ($R_P$) \cite{Farrar:1978xj} on the Lagrangian density, making the lightest
supersymmetric particle a stable dark-matter candidate by the same occasion. At the level of
renormalizable terms, $B$ and $L$ would then again appear as accidental symmetries of the
model. Nevertheless, if the MSSM is regarded as an effective field theory (EFT) at the electroweak (EW)
and SUSY scales, non-renormalizable operators acquire a legitimacy as markers of effects of
higher-energy (\textit{e.g.}\ a GUT or string completion), so that $R_P$-conserving $B$/$L$-violating operators
of dimension five could then develop and mediate proton decay \cite{Weinberg:1981wj,Sakai:1981pk,Allanach:2003eb}. 
Alternatively, $R_P$ could be sacrificed altogether, leading to
so-called $R_P$-violating (RpV) models \cite{Dreiner:1997uz,Barbier:2004ez}, with (renormalizable)
bilinear and/or trilinear either $B$- or $L$-violating terms in the superpotential. At this level, it is still
possible to impose $B$- or $L$-invariance on the Lagrangian density, or accept proton decay as a
phenomenological possibility.

These theoretical motivations, especially in the context of GUT models, have triggered extensive experimental interest in 
discovering $B$-violating decays of nucleons. Early experiments testing the law of $B$-conservation proposed by Weyl, 
Stueckelberg and Wigner \cite{Weyl:1929fm,Stueckelberg:1938zz,Wigner:1949zz} actually pre-date the Sakharov paper 
from 1967 \cite{Sakharov:1967dj}: a first experiment was performed in 1954 by Goldhaber, and also by Reines, Cowan, 
and Goldhaber \cite{Reines:1954pg}. See Table~I in Ref.~\cite{Gurr:1967pc} for a list of early experiments 1954-1964, as 
well as the work by Gurr \textit{et al.} in 1967. However, even the most recent results
\cite{Seidel:1988ut,Phillips:1989dp,Berger:1991fa,McGrew:1999nd,Kobayashi:2005pe,Regis:2012sn,Abe:2013lua,Abe:2014mwa,Miura:2016krn,TheSuper-Kamiokande:2017tit,Tanabashi:2018oca} 
have found no evidence for this phenomenon and place ever stronger bounds on individual proton or neutron decay channels. 
Recently, several new experiments \cite{An:2015jdp,Acciarri:2015uup,Abe:2018uyc} have been announced; they should be able 
to extend the current sensitivity considerably. See also the  recent overview in the introduction of  Ref.~\cite{Ellis:2019fwf}.

In R-parity conserving supersymmetry, dimension-5 baryon number violating operators have been considered 
\cite{Weinberg:1981wj}. 
However they involve external superpartners, which at low-energies  must be converted to SM 
particles, thus reverting to dimension-6 operators, with possibly more than one high-energy mass scale. A complete list of 
dimension-5 lepton- or baryon-number violating operators is given in Ref.~\cite{Allanach:2003eb}.

\begin{figure}[t]
\begin{center}
\includegraphics[height=3cm,width=7cm,angle=0]{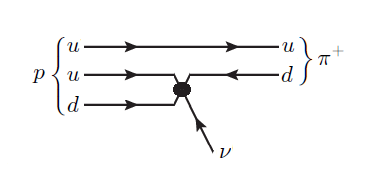}
\caption{Possible diagram for proton decay via an effective operator generated from the R-parity violating operators 
$U^cD^cD^c$ and $LQD^c$ in the superpotential.}
\label{fig:nuc-decay1}
\end{center}
\end{figure}

In this paper, we focus on nucleon decay from the RpV perspective, \textit{i.e.}\ with low-energy renormalizable
couplings and mediators relatively close to the EW scale. The superpotential of the $R_p$-conserving
MSSM is extended by the following  terms \cite{Weinberg:1981wj}:
\begin{eqnarray}
{W_{\not{R}_p}}&=&\mu_i H_u \cdot L_i+\frac{1}{2}\lambda_{ijk}L_i\cdot L_j
                                 (E^c)_k +\lambda'_{ijk}L_i\cdot  Q_j (D^c)_k\nonumber \\
 &&                           {     +\frac{1}{2}\lambda''_{ijk}
	\varepsilon_{\alpha\beta\gamma}(U^{c})^{\alpha}_i(D^{c})^{\beta}_j(D^{c})^{\gamma}_k,}
\label{eq:RpVSuperpotential}
\end{eqnarray}
where $Q$, $U^c$, $D^c$, $L$, $E^c$
denote the usual quark and lepton superfields, $\cdot$\ is the $SU(2)_L$ invariant antisymmetric product and
$\varepsilon_{\alpha\beta\gamma}$ is the 3-dimensional Levi-Civita symbol. The indices $i$, $j$, $k$
correspond to the three generations of flavor, while $\alpha$, $\beta$, $\gamma$ refer to the color
index. The parameters $\lam_{ijk}$ and $\lambda''_{ijk}$ satisfy the following conditions without
loss of generality: $\lambda_{ijk}=-\lambda_{jik}$, $\lambda''_{ijk}=-\lambda''_{ikj}$. The first three sets 
of terms of Eq.~(\ref{eq:RpVSuperpotential}) violate $L$ and the last one, $B$. The simultaneous
existence of $B$- and $L$-violating couplings opens up decay channels of nucleons into mesons 
and (anti)leptons, where squarks appear as typical mediators at tree-level. See 
Fig.~\ref{fig:nuc-decay1}, where we show the decay $p\to\pi^+\nu$ via an effective four-fermion 
interaction generated from the $U^cD^cD^c$ and $LQD^c$ operators. Such nucleon decays have 
received attention for a long time in the RpV MSSM \cite{Hall:1983id}; see \textit{e.g.}\ \cite{Barbier:2004ez,Dudas:2019gkj} for
summaries. Original studies focused on $(B$--$L)$-conserving processes \cite{Hinchliffe:1992ad}, then
$(B$+$L)$-conserving ones \cite{Vissani:1995hp,Bhattacharyya:1998dt}. Ref.~\cite{Smirnov:1996bg} 
observed that flavor flips associated to the charged weak interaction could be exploited to extend the limits to 
all flavor directions of the RpV couplings.\footnote{See also Refs.~\cite{Agashe:1995qm,Dreiner:1991pe} for the
effects of flavor flips on bounds on and also on signals of RpV, beyond proton decay.} For related cosmological 
bounds see for example \cite{Bouquet:1986mq,Campbell:1990fa,Dreiner:1992vm}. Beyond the `direct' nucleon 
decays mediated by a virtual squark exchange, slightly more complicated structures involving additional 
intermediate charginos and neutralinos were also considered
\cite{Hall:1983id,Zwirner:1984is,Carlson:1995ji,Hoang:1997kf,Bhattacharyya:1998bx}. In case such decays are 
kinematically allowed, these supersymmetric fermions \cite{Chang:1996sw}, or more exotic new particles 
\cite{Choi:1996nk,Choi:1998ak}, could also replace the lepton in the final state.  The case of a very light 
neutralino is still experimentally allowed  \cite{Choudhury:1999tn,Dreiner:2009ic,Dreiner:2009er,Dreiner:2011fp}.
and can also be searched for in rare meson decays in various experiments 
\cite{Dedes:2001zia,deVries:2015mfw,Dercks:2018eua,Dercks:2018wum,Dreiner:2020qbi}.

In the current paper, we attempt to update the status of the limits applying to the RpV couplings,
providing a more detailed attention to the low-energy form factors, about which the RpV literature
remains cursory, in general. We restrict ourselves to tree-level RpV contributions, since a full
one-loop matching would be much more involved. We also renounce a heuristic implementation of the
quark-flavor changes, as proposed in \textit{e.g.}\ Ref.~\cite{Smirnov:1996bg}, since we believe that 
such limits depend on the renormalization scheme, \textit{i.e.}\ on the formal definition of the tree-level 
couplings.\footnote{It is indeed possible to thus inflate the set of limits applying to individual (or pairs of) 
RpV couplings, but the actual bounds in fact constrain given directions in parameter space.} In the
following section, we introduce the EFT encoding nucleon decays and derive the matching conditions. We
also discuss the relevant low-energy hadronic matrix elements, referring to lattice evaluations, when
available, then comparing these results to those of a static bag model, which we employ in other
cases. Finally, in Sec.~3, we apply up-to-date experimental bounds to specific decay channels and
obtain limits on combinations of RpV couplings, before a short conclusion.

\section{Matching the RpV contributions on the \boldmath $\Delta B=1$ Hamiltonian}
\label{sec:matching}
In this section, we review the general framework that we employ to compute the nucleon decay widths in the context of the RpV MSSM.

\subsection{Low-energy EFT and QCD-running}
\label{sec:low-energy-eft}

The classification of operators involving only SM fields and satisfying the SM gauge symmetries
and violating $B$ was performed in \cite{Weinberg:1979sa,Wilczek:1979hc}. The operators of lowest
dimension that do not conserve $B$ are of dimension 6 and conserve $B-L$ \cite{Weinberg:1979sa}. Among
them, we will be more particularly interested in:
\begin{eqnarray}\label{eqn:dim6}
O^{(1)}_{mnpq}
  &=&\varepsilon_{\alpha\beta\gamma}\left[(\overline{d^c})_m^{\alpha}P_R(u)_n^{\beta}\right]\nonumber \\
  &&\cdot\left[(\overline{u^c})_p^{\gamma}
    P_L(e)_q-(\overline{d^c})_p^{\gamma}P_L(\nu)_q\right]\,,\\
  O^{(5)}_{mnpq}&=&\varepsilon_{\alpha\beta\gamma}\left[(\overline{d^c})_m^{\alpha}P_R(u)_n^{\beta}\right]
                  \left[(\overline{u^c})_p^{\gamma}P_R(e)_q\right]\,.\nonumber
\end{eqnarray}
$u$, $d$, $e$, $\nu$ correspond to the usual four-component spinors representing quarks and leptons,
with Latin index relating to flavor and Greek to color. $f^c$ ($f=u,d,e,\nu$) indicates
charge conjugation: $f^c=C\bar{f}^T$, with $C$ the charge-conjugation matrix. $P_{L,R}$ are chiral
projectors. We note that the fields are defined in the gauge-eigenstate basis, so that an additional CKM
rotation on (\textit{e.g.}) the down-type left-handed quarks and a PMNS rotation on the neutrinos should be
included in case we wish to work in the mass-eigenbasis.

\begin{figure*}[t]
\begin{center}
\includegraphics[height=3cm,width=17cm,angle=0]{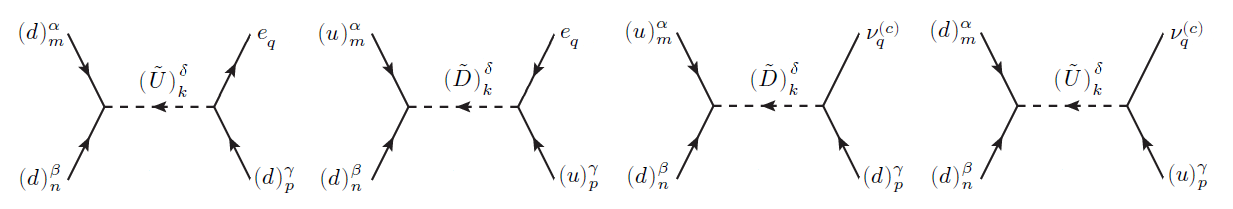}
\caption{Possible nucleon decays via the combination of couplings $\lam''_{kmn}$ and $\lam'_{qpk}$. These can be seen
as $t$- or $s$-channel processes.
\label{fig:possible-nucleon-decays}
}
\end{center}
\end{figure*}

Dimension-7 operators conserve $B+L$ \cite{Kobach:2016ami}. Through a Higgs vacuum expectation value (vev.) they produce
effective dimension 6 operators which violate the EW symmetry. We will encounter the following ones:
\begin{align}\label{eqn:dim7}
Q^{(1)}_{mnpq}&=\varepsilon_{\alpha\beta\gamma}\left[(\overline{d^c})_m^{\alpha}P_R(u)_n^{\beta}\right]
\left[(\overline{\nu})_qP_R(d)_p^{\gamma}\right]\,,\nonumber\\[1.5mm]
Q^{(2)}_{mnpq}&=\varepsilon_{\alpha\beta\gamma}\left[(\overline{d^c})_m^{\alpha}P_R(d)_n^{\beta}\right]
\left[(\overline{\nu})_qP_R(u)_p^{\gamma}\right]\,,\\[1.5mm]
Q^{(5)}_{mnpq}&=\varepsilon_{\alpha\beta\gamma}\left[(\overline{d^c})_m^{\alpha}P_R(d)_n^{\beta}\right]
\left[(\overline{e})_qP_L(d)_p^{\gamma}\right]\nonumber\,,\\[1.5mm]
Q^{(6)}_{mnpq}&=\varepsilon_{\alpha\beta\gamma}\left[(\overline{d^c})_m^{\alpha}P_R(d)_n^{\beta}\right]
\left[(\overline{e})_qP_R(d)_p^{\gamma}\right]\nonumber\,.
\end{align}
Such terms are produced in the RpV MSSM via particle mixing, either in the squark or in the lepton-higgsino-gaugino sectors.  
An EW-violating vev.\ is always needed to generate them. As long as the EW and SUSY scales are not resolutely in hierarchical 
ratio, \textit{i.e.} $M_{SUSY}/M_W$ is not too big, the associated suppression is not paramount. In fact, even the relic
of a dimension 8 operator will show up in tree-level matching, though involving two orders of mixing:
\begin{equation}
  R=\varepsilon_{\alpha\beta\gamma}\left[(\overline{d^c})_m^{\alpha}P_R(d)_n^{\beta}\right] \left[
    (\overline{\nu^c})_qP_L(u)_p^{\gamma}\right].
\end{equation}

In all the operators considered above, the lepton field can be replaced by an electroweakino field. In fact, due to the mixing 
appearing in the RpV context, the neutrinos and charged leptons could themselves be viewed as specific neutralino and chargino 
eigenstates. The resulting new operators could be genuine low-energy operators in the presence of \textit{e.g.}\ a light gaugino. 
If, on the contrary, the electroweakinos are very massive (as compared to the nucleon mass), these operators with external 
electroweakinos are simply a step in the direction of typically higher-dimensional low-energy operators, as considered in \textit{
e.g.}\ Ref.~\cite{Bhattacharyya:1998bx}. As long as no further quark (or gluon) lines are attached in this manner, the QCD aspects of 
the operators with external electroweakinos do not differ from those of operators with external leptons (up to momentum-dependent 
terms), so that the recipes discussed below continue to apply.

In the RpV MSSM, $B$-violating effects in nucleon decays are mediated by supersymmetric particles. At least the sfermions 
can be expected to be comparatively heavy with respect to the scale at which nucleon decay takes place. This means that, 
below the scale of the sfermions, we can summarize their impact in the $B$-violating processes by their contribution to the 
operators of Eqs.~(\ref{eqn:dim6}-\ref{eqn:dim7}) (where, technically, the operators of Eqs.~(\ref{eqn:dim7}) should be 
restored to their full EW-conserving version). This defines the effective Hamiltonian:
\begin{equation}
{\cal H}^{\text{eff}}=\sum_{\Omega=O,Q}C_{\Omega}(\mu_{R})\Omega(\mu_{R})\,,
\end{equation}
where $\mu_{R}$ denotes the renormalization scale. The Wilson coefficients $C_{\Omega}$ encode the 
short-distance effects and are obtained from integrating out the heavy fields. Large $\ln\frac{M_{\text{SUSY}}}{M_N}$ corrections, 
where $M_N$ denotes the nucleon mass, are expected to develop via radiative effects between the scale of the sfermions, 
$\mu_{R}=M_{\text{SUSY}}$, where the short-distance effects are defined, and the scale of the nucleon, $\mu_
{R}\approx M_N$, at which the operators mediate the hadronic process. The leading contributions to the anomalous
dimension have been studied in Ref.~\cite{Abbott:1980zj}.  Contrary to the case of GUTs, our high-energy boundary, the sfermion scale,  is expected to be 
comparatively close to the EW scale, so that we can neglect the EW-running and restrict ourselves to the sole QCD-running. 
Following Ref.~\cite{Abbott:1980zj}, all the operators then receive a simple scaling factor from the QCD corrections, which we can 
summarize as follows:
\begin{eqnarray}\label{eqn:running}
  C_{\Omega}(\mu_{R})&=&\eta_{\text{QCD}}\, C_{\Omega}(M_{\text{SUSY}})\,, \\[3mm]
  \eta_{\text{QCD}} &=&\left[\frac{\alpha_S(m_t)}{\alpha_S(M_{\text{SUSY}})}\right]^{2/\beta_0[6]}\cdot
                        \left[\frac{\alpha_S(m_b)}{\alpha_S(m_t)}
\right]^{2/\beta_0[5]}\nonumber\\&&\cdot\left[\frac{\alpha_S(\mu_{R})}{\alpha_S(m_b)}\right]^{2/\beta_0[4]}\,,\nonumber\\[3mm]
   \beta_0[N_F]&\equiv&11-\frac{2}{3}N_F\,, \nonumber
\end{eqnarray}
with $\alpha_S$ the running QCD coupling. The low-energy scale $\mu_{R}$ cannot be set
much below the charm mass, $m_c$, because of the perturbative description failing at low energy. Lattice calculations
\cite{Aoki:2017puj} employ $\mu_{R}=2$~GeV. Then, the problem is factorized in two
separate issues, the determination of the short-distance coefficients $C_{\Omega}(M_{\text{SUSY}})$ (or
matching to the high-energy model), which we perform in the next subsection,
and the evaluation of the hadronic matrix element, which we discuss in the subsequent subsections.


\subsection{Defining the Wilson coefficients}
In order to define the Wilson coefficients, we consider partonic scattering amplitudes both in the full RpV MSSM and in the 
EFT, and identify them at the SUSY scale (matching). See Fig.~\ref{fig:possible-nucleon-decays} for example interactions.

\paragraph{Feynman amplitude in the RpV MSSM --} These transition amplitudes can be easily written at
tree-level in terms of the couplings that are defined in the appendix.

An internal sup (up-squark) line mediates a transition amplitude with three external down-type quarks (for simplicity, we write 
fields in the amplitudes below, while they should be replaced by four-component spinors in practice):
\begin{eqnarray}
{\cal A}^{\text{RpV}}[d_m^{\alpha}d_n^{\beta}d_p^{\gamma}e_q]&=&\frac{i
\,\varepsilon_{\gamma\alpha\beta}}
                                                                 {m^2_{\tilde U_k}}(g_{R}^{Udd})_{kmn}\nonumber
  \\ &&\hspace{-1cm}\left\{(g_{L}^{Ud\chi})_{kpq}[(\overline{d^c})_m^{\alpha}
                                                                 P_R(d)_n^{\beta}][(\overline{e})_qP_L(d)_p^{\gamma}]\right.\nonumber\\
    &&\hspace{-1cm}\left.+(g_{R}^{Ud\chi})_{kpq}[(\overline{d^c})_m^{\alpha}P_R(d)_n^{\beta}][(\overline{e})_qP_R(d)_p^{\gamma}]\right\}\nonumber
                                                                \\ &&\hspace{-1cm}+[(m,\alpha)\leftrightarrow(n,\beta)\leftrightarrow(p,\gamma)]\,.
                                                                \label{eqn:ddde}
\end{eqnarray}
Here, $m_{\tilde U_k}$ denotes the sup mass of generation $k$, and the couplings $(g_{R}^{Udd})_{kmn}$ \textit{etc.} are defined in 
Appendix~\ref{ap:FR}.
The lepton spinor is one of the light states of the chargino-lepton system: $e_q=\chi^-_{q+2}$ (we
omit the $+2$ above, as well as the $+4$ for the neutrinos among the neutralino states later on). We see
that the result projects on the operators $Q^{(5)}_{mnpq}$ and $Q^{(6)}_{mnpq}$ of Eq.~(\ref{eqn:dim7}).

Similarly, with two entering up-type lines, one entering down-type line and a lepton, we have a diagram
with an internal sdown line:
\begin{eqnarray}
{\cal
  A}^{\text{RpV}}[u_m^{\alpha}d_n^{\beta}u_p^{\gamma}e_q]&=&-\frac{i\,\varepsilon_{\alpha\beta\gamma}}{m^2_{\tilde D_k}}(g_{R}^{uDd})_{mkn}
\nonumber \\
 &&\hspace{-1cm}\left\{(g_{L}^{Du\chi})_{kpq}[(\overline{u^c})_m^{\alpha}P_R(d)_n^{\beta}][(\overline{e^c})_qP_L(u)_p^{\gamma}]\right. \nonumber\\
  &&\hspace{-1cm}\left.+(g_{R}^{Du\chi})_{kpq}[(\overline{u^c})_m^{\alpha}P_R(d)_n^{\beta}][(\overline{e^c})_qP_R(u)_p^{\gamma}]\right\}\nonumber
\\ &&\hspace{-1cm} +[(m,\alpha)\leftrightarrow(p,\gamma)]\,.
\label{eqn:udue}
\end{eqnarray}
Here $m^2_{\tilde D_k}$ denotes the sdown mass and again the couplings $(g_{R}^{uDd})_{mkn}$ are given in Appendix~\ref{ap:FR}.
For two entering down-type lines and one entering up-type line plus a neutrino, we first have a diagram
with internal sdown line (the equation is somewhat abusive as we consider neutrino and antineutrino
simultaneously):
\begin{eqnarray}
{\cal
  A}^{\text{RpV}}[u_m^{\alpha}d_n^{\beta}d_p^{\gamma}\nu_q]&=&-\frac{i\,\varepsilon_{\alpha\beta\gamma}}
  {m^2_{\tilde D_k}}(g_{R}^{uDd})_{mkn}\nonumber\\
  && \hspace{-1cm} \left\{(g_{L}^{Dd\chi})_{kpq}[(\overline{u^c})_m^{\alpha}P_R(d)_n^{\beta}][(\overline{\nu^c})_qP_L(d)_p^{\gamma}]\right.\nonumber\\
  &&\hspace{-1cm}  \left.+(g_{R}^{Dd\chi})_{kpq}[(\overline{u^c})_m^{\alpha}P_R(d)_n^{\beta}][(\overline{\nu})_qP_R(d)_p^{\gamma}]\right\}\nonumber\\ 
   &&\hspace{-1cm}+[(n,\beta)\leftrightarrow(p,\gamma)]\,.
  \label{eqn:udde}
  \end{eqnarray}
Then, there is a diagram with internal sup line contributing to the same amplitude:
\begin{eqnarray}
{\cal
  A}^{\text{RpV}}[d_m^{\alpha}d_n^{\beta}u_p^{\gamma}\nu_q]&=&\frac{i\,\varepsilon_{\gamma\alpha\beta}}
                                                               {m^2_{\tilde U_k}}(g_{R}^{Udd})_{kmn}\nonumber \\
 &&\hspace{-1cm}\left\{(g_{L}^{Uu\chi})_{kpq}[(\overline{d^c})_m^{\alpha}P_R(d)_n^{\beta}][(\overline{\nu^c})_qP_L(u)_p^{\gamma}]\right.\nonumber\\
 &&\hspace{-1cm} \left.+(g_{R}^{Uu\chi})_{kpq}[(\overline{d^c})_m^{\alpha}P_R(d)_n^{\beta}][(\overline{\nu})_qP_R(u)_p^{\gamma}]\right\}
\nonumber \\
&&\hspace{-1cm}+[(m,\alpha)\leftrightarrow(n,\beta)]\,.
\label{eqn:ddue}
\end{eqnarray}

\paragraph{Amplitudes in the effective field theory --}
The corresponding amplitudes in the EFT read:
\begin{eqnarray}
{\cal A}^{\text{EFT}}[d_m^{\alpha}d_n^{\beta}d_p^{\gamma}e_q]&=&i\,\varepsilon_{\alpha\beta\gamma}\left\{(C_{Q_5})_{mnpq}\right. \nonumber\\
&&\hspace{-1cm}  [(\overline{d^c})_m^{\alpha}P_R(d)_n^{\beta}][(\overline{e})_qP_L(d)_p^{\gamma}]\nonumber\\
&&\hspace{-1cm}\left.+(C_{Q_6})_{mnpq}[(\overline{d^c})_m^{\alpha}P_R(d)_n^{\beta}][(\overline{e})_qP_R(d)_p^{\gamma}]\right\}\nonumber\\
                                                             &&\hspace{-1cm} +[(m,\alpha)\leftrightarrow(n,\beta)\leftrightarrow(p,\gamma)]\,,\label{eqn:dddeEFT}\\[1.5mm]
{\cal A}^{\text{EFT}}[u_m^{\alpha}d_n^{\beta}u_p^{\gamma}e_q]&=&-i\,\varepsilon_{\alpha\beta\gamma}\left\{(C_{O_1})_{nmpq}\right.\nonumber\\
&&\hspace{-1cm}  [(\overline{u^c})_m^{\alpha}P_R(d)_n^{\beta}][(\overline{e^c})_qP_L(u)_p^{\gamma}]\nonumber \\
 &&\hspace{-1cm}\left.+(C_{O_5})_{nmpq}[(\overline{u^c})_m^{\alpha}P_R(d)_n^{\beta}][(\overline{e^c})_qP_R(u)_p^{\gamma}]\right\}\nonumber
  \\
  && \hspace{-1cm}+[(m,\alpha)\leftrightarrow(p,\gamma)]\,,\label{eqn:udueEFT} 
\end{eqnarray}
\begin{eqnarray}
  {\cal  A}^{\text{EFT}}[u_m^{\alpha}d_n^{\beta}d_p^{\gamma}\nu_q]&=&-i\,\varepsilon_{\alpha\beta\gamma}\left\{(-C_{O_1})_{nmpq}\right. \nonumber\\
&& [(\overline{u^c})_m^{\alpha}P_R(d)_n^{\beta}][(\overline{\nu^c})_qP_L(d)_p^{\gamma}]\nonumber \\
&&\hspace{-1cm}\left.+(C_{Q_1})_{nmpq}[(\overline{u^c})_m^{\alpha}P_R(d)_n^{\beta}][(\overline{\nu})_qP_R(d)_p^{\gamma}]\right.\nonumber\\
 &&\hspace{-1cm}\left.-(C_{R})_{mnpq}[(\overline{d^c})_m^{\alpha}P_R(d)_n^{\beta}][(\overline{\nu^c})_q
                                                                     P_L(u)_p^{\gamma}]\right.\nonumber\\
&&\hspace{-1cm}\left.-(C_{Q_2})_{mnpq}[(\overline{d^c})_m^{\alpha}P_R(d)_n^{\beta}][(\overline{\nu})_qP_R(u)_p^{\gamma}]\nonumber\right\}\nonumber
  \\ &&\hspace{-1cm}+[(n,\beta)\leftrightarrow(p,\gamma)]\,.\label{eqn:uddeEFT}
\end{eqnarray}

\paragraph{Matching --}
Identifying Eqs.~(\ref{eqn:dddeEFT}-\ref{eqn:uddeEFT}) with Eqs.~(\ref{eqn:ddde}-\ref{eqn:ddue}), we obtain:
\begin{eqnarray}
(C_{O_1})_{nmpq}&=&\frac{1}{m^2_{\tilde D_k}}(g_{R}^{uDd})_{mkn}(g_{L}^{Du\chi})_{kpq}\,,\nonumber \\
(C_{O_5})_{nmpq}&=&\frac{1}{m^2_{\tilde D_k}}(g_{R}^{uDd})_{mkn}(g_{R}^{Du\chi})_{kpq}\,,\nonumber\\
(C_{Q_1})_{nmpq}&=&\frac{1}{m^2_{\tilde D_k}}(g_{R}^{uDd})_{mkn}(g_{R}^{Dd\chi})_{kpq}\,, \nonumber\\
(C_{Q_2})_{mnpq}&=&\frac{1}{m^2_{\tilde U_k}}(g_{R}^{Udd})_{kmn}(g_{R}^{Uu\chi})_{kpq}\,,\\
  (C_{Q_5})_{mnpq}&=&\frac{1}{m^2_{\tilde U_k}}(g_{R}^{Udd})_{kmn}(g_{L}^{Ud\chi})_{kpq}\,,\nonumber\\
  (C_{Q_6})_{mnpq}&=&\frac{1}{m^2_{\tilde U_k}}(g_{R}^{Udd})_{kmn}(g_{R}^{Ud\chi})_{kpq}\,,\nonumber\\
  (C_{R})_{mnpq}&=&\frac{1}{m^2_{\tilde U_k}}(g_{R}^{Udd})_{kmn}(g_{L}^{Uu\chi})_{kpq}\,. \nonumber
\end{eqnarray}
The contribution to $O_1$ is mediated directly by $\lam''$ and $\lam'$ couplings. The contribution
to $O_5$ is in fact of dimension 8: it involves a Higgs vev.\ from squark mixing and a second from
chargino mixing. It is thus generated from the $\mu_{i}$ terms and receives additional mixing
suppression. The contributions to $Q_1$ and $Q_6$ can be generated from a $\lambda'$ coupling, in which
case the Higgs vev.\ is provided by squark mixing, or from $\mu_{i}$, in which case the vev.\ comes from
gaugino-higgsino mixing. The contribution to $Q_2$ is essentially mediated by mixing of the lepton with
the gauginos. The contributions to $Q_5$ and $R$ are of the same order as that to $O_5$, \textit{i.e.}\ depend on
secondary mixing of the leptons with the charginos/neutralinos. This counting is changed if we replace
the external leptons by electroweakino states, as we see in Sec.~\ref{sec:bounds}.

\subsection{Low-energy Operators}
So far, we have kept generic flavor indices. However, assuming that valence quarks determine nucleon decays, we can restrict ourselves to the three light quark flavors (as well as the two lighter charged leptons), hence to a smaller set of operators:
\begin{eqnarray}\label{eq:lowop}
\mathcal{O}_1^e&=&\varepsilon_{\alpha\beta\gamma}[(\overline{d^c})^{\alpha}P_Ru^{\beta}][(\overline{u^c})^{\gamma}P_Le] \,,\nonumber\\
\mathcal{O}_1^{\nu}&=&\varepsilon_{\alpha\beta\gamma}[(\overline{d^c})^{\alpha}P_Ru^{\beta}][(\overline{d^c})^{\gamma}P_L\nu]\,,\nonumber\\
  \mathcal{O}_5^e&=&\varepsilon_{\alpha\beta\gamma}[(\overline{d^c})^{\alpha}P_Ru^{\beta}][(\overline{u^c})^{\gamma}P_Re] \,,\nonumber\\
  \mathcal{Q}_1^{\nu}&=&\varepsilon_{\alpha\beta\gamma}[(\overline{d^c})^{\alpha}P_Ru^{\beta}][(\overline{d^c})^{\gamma}P_R\nu^c]\,,\nonumber\\
  &&\nonumber \\[-1.6mm]
\hat{\mathcal{O}}_1^e&=&\varepsilon_{\alpha\beta\gamma}[(\overline{s^c})^{\alpha}P_Ru^{\beta}][(\overline{u^c})^{\gamma}P_Le]\,,\nonumber\\
  \hat{\mathcal{O}}_1^{\nu}&=&\varepsilon_{\alpha\beta\gamma}[(\overline{s^c})^{\alpha}P_Ru^{\beta}][(\overline{d^c})^{\gamma}P_L\nu]\,,\nonumber\\
\hat{\mathcal{O}}_5^e&=&\varepsilon_{\alpha\beta\gamma}[(\overline{s^c})^{\alpha}P_Ru^{\beta}][(\overline{u^c})^{\gamma}P_Re]\,, \\
\hat{\mathcal{O}}'^{\nu}_1&=&\varepsilon_{\alpha\beta\gamma}[(\overline{d^c})^{\alpha}P_Ru^{\beta}][(\overline{s^c})^{\gamma}P_L\nu]\,,\nonumber\\
  \hat{\mathcal{Q}}_1^{\nu}&=&\varepsilon_{\alpha\beta\gamma}[(\overline{s^c})^{\alpha}P_Ru^{\beta}][(\overline{d^c})^{\gamma}P_R\nu^c]\,, \nonumber\\
 \hat{\mathcal{Q}}'^{\nu}_1&=&\varepsilon_{\alpha\beta\gamma}[(\overline{d^c})^{\alpha}P_Ru^{\beta}][(\overline{s^c})^{\gamma}P_R\nu^c]\,,\nonumber\\
\hat{\mathcal{Q}}_2^{\nu}&=&\varepsilon_{\alpha\beta\gamma}[(\overline{s^c})^{\alpha}P_Rd^{\beta}][(\overline{u^c})^{\gamma}P_R\nu^c]\,,\nonumber \\
 \hat{\mathcal{R}}^{\nu}&=&\varepsilon_{\alpha\beta\gamma}\left[(\overline{s^c})^{\alpha}P_Rd^{\beta}\right]\left[(\overline{u^c})^{\gamma}P_L\nu\right]\,,                 \nonumber\\
  \hat{\mathcal{Q}}_5^{e}&=&\varepsilon_{\alpha\beta\gamma}[(\overline{s^c})^{\alpha}P_Rd^{\beta}][(\overline{d^c})^{\gamma}P_Le^c]\,,\nonumber \\
\hat{\mathcal{Q}}_6^e&=&\varepsilon_{\alpha\beta\gamma}[(\overline{s^c})^{\alpha}P_Rd^{\beta}][(\overline{d^c})^{\gamma}P_Re^c] \nonumber \,.
\end{eqnarray}
Operators with $\hat{}$ (lower set) induce $\Delta S=1$, while those without preserve strangeness. We
neglect operators with $\Delta S\geq2$. The lepton should be understood as generic, $e=e,\mu$ and
$\nu=\nu_{e,\mu,\tau}$. We note that when twice the same quark is contracted in a scalar product, \textit{e.g.}\
$\varepsilon_{\alpha\beta\gamma}[(\overline{d^c})^{\alpha}P_{L,R}d^{\beta}]$, then the operator is
identically $0$.

It is then straightforward to identify the Wilson coefficients of the operators of Eq.~(\ref{eq:lowop})
with the low-energy coefficients of the original operators, Eqs.~(\ref{eqn:dim6}-\ref{eqn:dim7}):
\begin{eqnarray}
  C[\mathcal{O}^e_1]&=&(C_{O_1})_{111e}\,,\nonumber \\
  C[\hat{\mathcal{O}}^e_1]&=&(C_{O_1})_{211e}\,, \nonumber\\
 C[\mathcal{O}^{\nu}_1]&=&-V^{\text{CKM}}_{r1}(C_{O_1})_{11r\nu}\,, \nonumber \\
 C[\hat{\mathcal{O}}^{\nu}_1]&=&-V^{\text{CKM}}_{r1}(C_{O_1})_{21r\nu} \,,\nonumber \\
  C[\hat{\mathcal{O}}'^{\nu}_1]&=&-V^{\text{CKM}}_{r2}(C_{O_1})_{11r\nu}\,,\nonumber
 \end{eqnarray}
\begin{eqnarray}
 C[\mathcal{O}^e_5]&=&(C_{O_5})_{111e}\,,\nonumber \\
  C[\hat{\mathcal{O}}^e_5]&=&(C_{O_5})_{211e} \,, \\
   C[\mathcal{Q}^{\nu}_1]&=&(C_{Q_1})_{111\nu}\,, \nonumber \\
   C[\hat{\mathcal{Q}}^{\nu}_1]&=&(C_{Q_1})_{211\nu}\,, \nonumber \\
  C[\hat{\mathcal{Q}}'^{\nu}_1]&=&(C_{Q_1})_{112\nu}\,,\nonumber\\
  C[\hat{\mathcal{Q}}^{\nu}_2]&=&(C_{Q_2})_{211\nu}-(C_{Q_2})_{121\nu}\,,\nonumber \\
  C[\hat{\mathcal{R}}^{\nu}]&=&(C_{R})_{211\nu}-(C_{R})_{121\nu}\,,\nonumber\\
 C[\hat{\mathcal{Q}}^e_5]&=&V^{\text{CKM}}_{r1}\left[(C_{Q_5})_{21re}-(C_{Q_5})_{12re}\right]\,,\nonumber\\
  C[\hat{\mathcal{Q}}^e_6]&=&(C_{Q_6})_{211e}-(C_{Q_6})_{121e}\,. \nonumber
\end{eqnarray}
Here, we have chosen to define the couplings such that the CKM rotation $V^{\text{CKM}}$ is fully carried by the down-quark sector
\cite{Allanach:1999ic}.


\subsection{Hadronic Matrix Elements}
In order to connect the parton level $B$-violating EFT to actual nucleon decays, it is necessary to evaluate the
operators of Eq.~(\ref{eq:lowop}) between the nucleon and its hadronic decay products, \textit{e.g.}\
pions or kaons. Non-perturbative methods are needed to perform this step. Among the models employed in the
1980's, a first class
\cite{Jarlskog:1978uu,Machacek:1979tx,Goldman:1980ah,Kane:1980bn,Gavela:1981cf,Salati:1982bw} would
consider non-relativistic partons and exploit the $SU(6)$ flavor-spin symmetry. An alternative approach
is that of the bag models
\cite{Din:1979bz,Donoghue:1979pr,Golowich:1980ne,Donoghue:1982jm,Wakano:1982sk,Okazaki:1982eh}, where
partonic quarks are now  relativistic. Major conceptual difficulties in these descriptions appear in
association with \textit{e.g.}\ the treatment of a relativistic pion in the final state or the impact of the
three-quark fusion process \cite{Donoghue:1982jm,Berezinsky:1981qb}. Partial conservation of axial
vector currents was also employed for the estimate
\cite{Berezinsky:1981qb,Tomozawa:1980rc,DeLabastida:1981mp,Wise:1980ch}. The formulation of a chiral
model for baryon interaction \cite{Claudson:1981gh,Isgur:1982iz,Chadha:1983sj,Kaymakcalan:1983uc}
allowed to derive relations between the decay rates mediated by dimension 6 operators and low-energy
constants (LEC). How far the validity of the chiral model extends is not completely clear. Predictions
are (very) roughly in agreement among these calculations.

\subsubsection{Lattice Evaluation}

Lattice approaches to nucleon decays were also considered early on and have continued up to this day
\cite{Gavela:1988cp,Aoki:1999tw,Tsutsui:2004qc,Aoki:2006ib,Braun:2008ur,Aoki:2013yxa,Aoki:2017puj}. Corresponding
calculations focus on dimension 6 operators and nucleon ($N$) decays into one pseudoscalar meson ($\Pi$)
and one (anti)lepton ($L^{(c)}$): $N\to\Pi+L^{(c)}$. The matrix
elements can be represented by the following form-factors:
\begin{eqnarray}
\left<\Pi(p-\ell)\right|\Omega_{GH}\left|N(p)\right>&=&P_H\left\{W_{0,[\Omega_{GH}]}^{N\to\Pi}(\ell^2)\right.  \label{eq:formfac}\\
  &&\hspace{-1cm}\left.-\frac{i
     \ell\!\!/}{M_N}W_{1,[\Omega_{GH}]}^{N\to\Pi}(\ell^2)\right\}u_N(p)\,,\nonumber \\
\Omega_{GH}\equiv\varepsilon_{\alpha\beta\gamma}[(\overline{q^c})^{\alpha}_i&&\hspace{-0.5cm}P_G(q)^{\beta}_j][P_H(q)^{\gamma}_k]\,.
\nonumber
\end{eqnarray}
Here, $q=u,d,s$; $G,H\in\{L,R\}$ are the indices of the chiral projectors; $p$ and $\ell$ denote the
four-momenta of the nucleon and (anti)lepton, respectively; $u_N(p)$ represents the spinor associated with the nucleon,
$M_N$, its mass, and $W_{(0,1),[\Omega_{GH}]}^{N\to \Pi}$ correspond to the form factors, which depend on the momentum 
transfer squared, $\ell^2$, between the nucleon and the meson. For commodity, we will write
the left-hand side of Eq.~(\ref{eq:formfac}) in the abbreviated form
$\left<\Pi\right|(q_iq_j)_G(q_k)_H\left|N\right>$ below. If $q_i=q_j$, then the operator is identically
0. Parity-invariance of the strong interaction results in identities under $(L,R)\leftrightarrow(R,L)$
or $(L,L)\leftrightarrow(R,R)$ exchanges. The assumption of isospin invariance $u\leftrightarrow d$
produces further relations between different initial or final states: $p\leftrightarrow -n$,
$\pi^+\leftrightarrow\pi^-$, $\pi^0\to-\pi^0$, $\eta\to\eta$, $K^+\leftrightarrow K^0$,
$\rho^+\leftrightarrow\rho^-$, $\rho^0\to-\rho^0$, $\omega\rightarrow\omega$,
$K^{*+}\leftrightarrow K^{*0}$.

The lattice results for all the relevant form factors $W_{0,[\Omega]}^{N\to\Pi}$ (including the renormalization
scheme conversion),
are presented in Table~4 of Ref.~\cite{Aoki:2017puj}. The $W_{1,[\Omega]}^{N\to\Pi}$ corrections are evidently
suppressed for electrons and neutrinos in the final state, but have been considered in the case of a
muon (see Table~5 of the previous reference). We then obtain a complete list of matrix elements for the
operators of Eq.~(\ref{eq:lowop}) for the transitions of $N\to\Pi+L^{(c)}$ type:
\begin{eqnarray}
  \left<\Pi(p-\ell),L^{(c)}\right|\Omega_{G,H}\left|N(p)\phantom{L^{(c)}\hspace{-0.6cm}}\right>&=&\overline{w}_L(l)P_H\cdot
  \label{eq:mat-elem-nuc-mes-lep}
  \\
&&\hspace{-3cm}  
\cdot\left\{W_{0,[\Omega]}^{GH}(\ell^2)-\frac{i
   \ell\!\!\!/}{M_N}W_{1,[\Omega]}^{GH}(\ell^2)\right\}u_N(p)\,; \textcolor{white}{.} \nonumber 
\end{eqnarray}
$w_L=u_L,v_L^c$ denotes the lepton spinor.
From there, it is straightforward to derive the decay amplitudes and decay widths: 
\begin{widetext}
\begin{eqnarray}
{\cal A}[N\to\Pi+L^{(c)}]&=& i\sum_{\Omega}C[\Omega_{GH}]\left<\Pi(p-\ell),L^{(c)}\right|\Omega_{GH}\left|N(p)\phantom{L^{(c)}\hspace{-0.6cm}}\right>\\
                     &=&i\overline{w}_L(l)P_H \sum_{\Omega}C[\Omega_{GH}]W^{N\to\Pi}_{[\Omega_{GH}]}u_N(p)\,,\nonumber\\
  \Gamma[N\to\Pi+L^{(c)}]&=&\frac{1}{16\pi M_N}\sqrt{1-2\frac{M_{\Pi}^2+M_{L}^2}{M_N^2}+\left(\frac{M_{\Pi}^2-M_{L}^2}{M_N^2}\right)^2}
  \left[\frac{1}{2}\sum_{\text{pol.}}\left|{\cal A}[N\to\Pi+L^{(c)}]\right|^2\right]\\
                     &\simeq&\frac{M_N}{32\pi}\left(1-\frac{M_{\Pi}^2}{M_N^2}\right)^{2}
                                          \sum_{\Omega,\Omega'}(C[\Omega_{GH}]W^{N\to\Pi}_{[\Omega_{GH}]})^*
                                          (C[\Omega'_{G'H'}]W^{N\to\Pi}_{[\Omega'_{G'H'}]})\delta_{HH'}\,,\nonumber
\end{eqnarray}
\end{widetext}
with $\Omega$, $\Omega'$ scanning the list of Eq.~(\ref{eq:lowop}), $\displaystyle \sum_{\text{pol.}}$ representing 
the sum over spinor polarizations and $M_X$ denoting the mass of particle $X$. We have neglected the lepton mass 
in the last line. $W$ reduces to $W_0$ most of the time but includes the $W_1$ correction for muons in the final state.

\subsubsection{Non-Lattice Evaluation}

If we want to consider other decay channels, \textit{e.g.}\ involving vector mesons, we need to turn to other
means of estimates of the hadronic matrix elements, since corresponding lattice results are not
available. However, this means that the uncertainties will be significantly larger for these
additional channels. The lattice results are presented with comparatively small uncertainties
$\sim O(10\%)$ \cite{Aoki:2017puj}. This latter reference already mentions a factor $2$ to $3$
difference with the proton decays evaluated from LEC, showing that the precision drops considerably when
resorting to non-lattice methods. In practice, we consider the static bag model described in the
appendix, from which we expect results of strictly qualitative value. On the other hand, considering the
relatively large mass of the vector mesons, the static bag model should be performing almost in its
regime of validity. We first compared the results of the bag model with those obtained in
Ref.~\cite{Donoghue:1979pr}, which we essentially follow, and sensibly recovered the results of this paper
for the operator considered in this reference. Then, we consider the bag predictions for the $N\to\Pi+L$
channels in order to compare with the lattice results and assess the reliability of this approach. The
results are displayed in Table~\ref{tab:hadrmatpseudo} for a proton radius of $5~\text{GeV}^{-1}$ and
smaller meson radii (pion: $3.3~\text{GeV}^{-1}$, kaon: $2.8~\text{GeV}^{-1}$, $\eta$:
$4.7~\text{GeV}^{-1}$). In the case of `narrow' mesons (pions, kaons), the results of the bag model are
one order of magnitude below the central values originating from the lattice calculation, and somewhat
closer for `broad' mesons ($\eta$, a factor $\sim2$). The difference in the case of the narrow mesons
can be somewhat reduced if one employs the same bag radius of the proton for the mesons as well ---
essentially because the overlap between proton and meson wave-functions is larger --- and these values
could be further tuned by varying the chosen proton radius (which we do not attempt). We thus observe
that the bag model is not reliable on a quantitative basis. However, the corresponding results seem to
always underestimate the hadronic matrix elements, so that the bounds that one obtains when using these
values are conservative.
\begin{table*}
\begin{center}
\begin{tabular}{|c||c|c||c|c||}
\hline
& $W_0[p\to\pi^0]$ & $W_0[p\to\pi^0]$ & $W_0[p\to\eta]$ & $W_0[p\to\eta]$ \\
$(du)_G(u)_H$   & (GeV)$^2$, lat.& (GeV)$^2$, bag & (GeV)$^2$, lat.& (GeV)$^2$, bag    \\     
\hline\hline
$LL$, $RR$      & $0.134$        & $0.015$        & $0.113$        & $0.074$       \\
\hline
$LR$, $RL$      & $-0.131$       & $-0.015$       & $0.006$        & $0.005$       \\
\hline
\end{tabular}
		
\begin{tabular}{|c||c|c||}
\hline
& $W_0[p\to\pi^+]$ & $W_0[p\to\pi^+]$ \\
$(du)_G(d)_H$   & (GeV)$^2$, lat.& (GeV)$^2$, bag     \\     
\hline\hline
$LL$, $RR$      & $0.189$        & $0.022$        \\
\hline
$LR$, $RL$      & $-0.186$       & $-0.018$       \\
\hline
\end{tabular}
		
\begin{tabular}{|c||c|c||}
\hline
& $W_0[p\to K^0]$ & $W_0[p\to K^0]$ \\
$(su)_G(u)_H$   & (GeV)$^2$, lat.& (GeV)$^2$, bag    \\     
\hline\hline
$LL$, $RR$      & $0.057$        & $0.010$        \\
\hline
$LR$, $RL$      & $0.103$        & $0.012$        \\
\hline
\end{tabular}
		
\begin{tabular}{|c||c|c||c|c||c|c||}
\hline
$W_0[p\to K^+]$ & $(su)_G(d)_H$    & $(su)_G(d)_H$    & $(du)_G(s)_H$    & $(du)_G(s)_H$    & $(sd)_G(u)_H$    & $(sd)_G(u)_H$     \\
& (GeV)$^2$, lat.& (GeV)$^2$, bag  & (GeV)$^2$, lat.  & (GeV)$^2$, bag   & (GeV)$^2$, lat.  & (GeV)$^2$, bag        \\     
\hline\hline
$LL$, $RR$    & $0.041$          & $0.022$          & $0.139$          & $0.014$          & $-0.098$         & $-0.012$           \\
\hline
$LR$, $RL$    & $-0.049$         & $-0.022$         & $-0.134$         & $-0.014$         & $-0.054$         & $-0.010$           \\
\hline
\end{tabular}
\end{center}
\caption{Hadronic matrix elements for proton to pseudoscalar meson decays from lattice computations. 
Here the entries such as $(du)_G(u)_H$ denote the operator $\Omega_{G,H}$ of Eq.~(\ref{eq:mat-elem-nuc-mes-lep}).
We have not included the subscripts $\Omega_{G,H}$ on $W_0$ but just indicate the decay mode in brackets.}
\label{tab:hadrmatpseudo} 
\end{table*}

\begin{table*}
\begin{center}
\begin{tabular}{|c||c||c||c|c||c||}
\hline
& $W_0[p\to\rho^0]$ & $W_0[p\to\omega^0]$ & \null &                  &$W_0[p\to\rho^+]$ \\
$(du)_G(u)_H$  & (GeV)$^2$, bag  & (GeV)$^2$, bag    & \null & $(du)_G(d)_H$    &(GeV)$^2$, bag  \\     
\hline\hline
$LL$, $RR$     & $0.062$         & $-0.041$          & \null &$LL$, $RR$        & $0.088$        \\
\hline
$LR$, $RL$     & $-0.032$        & $0.041$           & \null &$LR$, $RL$        & $-0.048$       \\
\hline
\end{tabular}
	
\begin{tabular}{|c||c||c|c||c||c||c||}
\hline
& $W_0[p\to K^{*0}]$ & \null & $W_0[p\to K^{*+}]$ & $(su)_G(d)_H$    & $(du)_G(s)_H$    & $(sd)_G(u)_H$    \\
$(su)_G(u)_H$  & (GeV)$^2$, bag   & \null &                  & (GeV)$^2$, bag   & (GeV)$^2$, bag   & (GeV)$^2$, bag   \\     
\hline\hline
$LL$, $RR$     & $-0.005$         & \null & $LL$, $RR$       & $0.013$          & $0.022$          & $-0.008$         \\
\hline
$LR$, $RL$     & $-0.006$         & \null & $LR$, $RL$       & $0.009$          & $-0.022$         & $-0.002$          \\
\hline
\end{tabular}
\end{center}

\caption{Hadronic matrix elements for proton to vector meson decays using the bag model. Otherwise the notation is as in Table~\ref{tab:hadrmatpseudo}.}
\label{tab:hadrmatvect} 
\end{table*}

The hadronic matrix elements for the transitions into vector mesons are shown in Table~\ref{tab:hadrmatvect} with 
meson radii $4.8~\text{GeV}^{-1}$ (rho, omega) and $3.0~\text{GeV}^{-1}$ ($K^*$). There, we define the form factors as:
\begin{widetext}
\begin{eqnarray}
\left<V(p-\ell)\right|\Omega_{GH}\left|N(p)\right>&=&W^{N\to V}_{[\Omega_{GH}]}(\ell^2)\,P_H\gamma^{\mu}u_N(p)\varepsilon^{V*}_\mu(p-\ell)\,,
\end{eqnarray}
where $\varepsilon^{V*}_\mu$ corresponds to the polarization vector of the vector meson. Including the lepton spinor $L^{(c)}$ in
the matrix element, the decay amplitude and widths then read:
\begin{align}
&{\cal A}[N\to V+L^{(c)}]=i\sum_{\Omega}C[\Omega_{GH}]\left<V(p-\ell),L^{(c)}\right|\Omega_{GH}\left|N(p)\phantom{L^{(c)}\hspace{-0.6cm}}\right>\\
&\null\hspace{2.7cm}=i\sum_{\Omega}C[\Omega_{GH}]W^{N\to V}_{[\Omega_{GH}]}\,\overline{w}_L(l)P_H\gamma^{\mu}u_N(p)\varepsilon^{V*}_\mu(p-\ell)\,,\nonumber\\
&\Gamma[N\to V+L^{(c)}]=\frac{1}{16\pi M_N}\sqrt{1-2\frac{M_{V}^2+M_{L}^2}{M_N^2}+\left(\frac{M_{V}^2-M_{L}^2}{M_N^2}\right)^2}\left[\frac{1}{2}\sum_{\text{pol.}}\left|{\cal A}[N\to V+L^{(c)}]\right|^2\right]\\
&\null\hspace{2.6cm}\simeq\frac{M_N^3}{64\pi M_V^2}\left(1-\frac{M_{V}^2}{M_N^2}\right)^{2}\left(1+2\frac{M_V^2}{M_N^2}\right)\sum_{\Omega,\Omega'}
(C_{\Omega}^{GH}W^{N\to V}_{[\Omega_{GH}]})^*(C[\Omega'_{G'H}]W^{N\to V}_{[\Omega'_{G'H}]})\,.\nonumber
\end{align}
\end{widetext}
We have again neglected the lepton mass in the last expression.
Given that the matrix elements obtained with the bag model are suppressed as compared to those 
derived on the lattice, the limits applying to nucleon to vector meson transitions will always prove 
subleading as compared to those of the nucleon to pseudoscalar meson channels. As the latter are 
more reliable, this situation is desirable so that derived bounds are conservative. On the other hand, 
we note that in the bag model, the branching ratios of the nucleon to vector meson transitions are 
larger than those of the nucleon decays into pseudoscalar mesons. Consequently, a more precise
determination of the hadronic form factors for vector channels could eventually lead to competitive 
results, at least in certain directions of the $B$-violating operator basis.

The matrix elements for the neutron decays can be obtained from those of the proton channels by exploiting the isospin symmetry:
\begin{eqnarray}
 \left<\pi^-\right|(du)_G(u)_H\left|n\phantom{\pi^-}\hspace{-0.4cm}\right>&=&\left<\pi^+\right|(du)_G(d)_H\left|p\phantom{\pi^-}\hspace{-0.4cm}\right>\,,
  \nonumber \\
  \left<\pi^0\right|(du)_G(d)_H\left|n\phantom{\pi^-}\hspace{-0.4cm}\right>&=&-\left<\pi^0\right|(du)_G(u)_H\left|p\phantom{\pi^-}\hspace{-0.4cm}\right>\,, \nonumber\\
 \left<\eta\right|(du)_G(d)_H\left|n\phantom{\pi^-}\hspace{-0.4cm}\right>&=&\left<\eta\right|(du)_G(u)_H\left|p\phantom{\pi^-}\hspace{-0.4cm}\right>\,, \nonumber  \\
 \left<K^+\right|(sd)_G(d)_H\left|n\phantom{\pi^-}\hspace{-0.4cm}\right>&=&-\left<K^0\right|(su)_G(u)_H\left|p\phantom{\pi^-}\hspace{-0.4cm}\right>\,, \nonumber \\ 
 \left<K^0\right|(du)_G(s)_H\left|n\phantom{\pi^-}\hspace{-0.4cm}\right>&=&\left<K^+\right|(du)_G(s)_H\left|p\phantom{\pi^-}\hspace{-0.4cm}\right>\,, \nonumber \\
 \left<K^0\right|(su)_G(d)_H\left|n\phantom{\pi^-}\hspace{-0.4cm}\right>&=&-\left<K^+\right|(sd)_G(u)_H\left|p\phantom{\pi^-}\hspace{-0.4cm}\right>\,, \nonumber \\ 
\left<K^0\right|(sd)_G(u)_H\left|n\phantom{\pi^-}\hspace{-0.4cm}\right>&=&-\left<K^+\right|(su)_G(d)_H\left|p\phantom{\pi^-}\hspace{-0.4cm}\right>\,, \\
 \left<\rho^-\right|(du)_G(u)_H\left|n\phantom{\pi^-}\hspace{-0.4cm}\right>&=&\left<\rho^+\right|(du)_G(d)_H\left|p\phantom{\pi^-}\hspace{-0.4cm}\right>\,, \nonumber \\ 
 \left<\rho^0\right|(du)_G(d)_H\left|n\phantom{\pi^-}\hspace{-0.4cm}\right>&=&-\left<\rho^0\right|(du)_G(u)_H\left|p\phantom{\pi^-}\hspace{-0.4cm}\right>\,, \nonumber\\
\left<\omega\right|(du)_G(d)_H\left|n\phantom{\pi^-}\hspace{-0.4cm}\right>&=&\left<\omega\right|(du)_G(u)_H\left|p\phantom{\pi^-}\hspace{-0.4cm}\right>\,,  \nonumber\\
 \left<K^{*+}\right|(sd)_G(d)_H\left|n\phantom{\pi^-}\hspace{-0.4cm}\right>&=&-\left<K^{*0}\right|(su)_G(u)_H\left|p\phantom{\pi^-}\hspace{-0.4cm}\right>\,, \nonumber 
\end{eqnarray}
\begin{eqnarray}
\left<K^{*0}\right|(du)_G(s)_H\left|n\phantom{\pi^-}\hspace{-0.4cm}\right>&=&\left<K^{*+}\right|(du)_G(s)_H\left|p\phantom{\pi^-}\hspace{-0.4cm}\right>\,, \nonumber\\
\left<K^{*0}\right|(su)_G(d)_H\left|n\phantom{\pi^-}\hspace{-0.4cm}\right>&=&-\left<K^{*+}\right|(sd)_G(u)_H\left|p\phantom{\pi^-}\hspace{-0.4cm}\right>\,, \nonumber \\ 
\left<K^{*0}\right|(sd)_G(u)_H\left|n\phantom{\pi^-}\hspace{-0.4cm}\right>&=&-\left<K^{*+}\right|(su)_G(d)_H\left|p\phantom{\pi^-}\hspace{-0.4cm}\right>\,. \nonumber
\end{eqnarray}


\subsection{Summary}
We can summarize the calculation of the nucleon decay widths into a meson and a(n anti)lepton with the
following equations:
\begin{eqnarray}
\tilde{\Gamma}[p\to(\pi^0,\eta,\rho^0,\omega^0)+e^+]&=&\left|W^{N\to M}_{\mathcal{O}_1^e}C[\mathcal{O}_1^e]\right|^2
\nonumber\\ 
&&+\left|W^{N\to M}_{\mathcal{O}^e_5}C[\mathcal{O}^e_5]\right|^2\nonumber\\
&=&\tilde{\Gamma}[n\to(\pi^-,\rho^-)+e^+]\,,\nonumber\\
\tilde{\Gamma}[p\to(\pi^+,\rho^+)+\nu^{(c)}]&=&\left|W^{N\to M}_{\mathcal{O}_1^{\nu}}C[\mathcal{O}_1^{\nu}]\right|^2\nonumber\\
&&+\left|W^{N\to M}_{\mathcal{Q}_1^{\nu}}C[\mathcal{Q}_1^{\nu}]\right|^2\nonumber\\
&=&\tilde{\Gamma}[n\to(\pi^0,\eta,\rho^0,\omega^0)+\nu^{(c)}]\,,\nonumber\\
\tilde{\Gamma}[p\to K^{0(*)}+e^+]&=&\left|W^{N\to M}_{\hat{\mathcal{O}}_1^{e}}C[\hat{\mathcal{O}}_1^{e}]\right|^2\nonumber\\
&&+\left|W^{N\to M}_{\hat{\mathcal{O}}^e_5}C[\hat{\mathcal{O}}^e_5]\right|^2\,,\\
\tilde{\Gamma}[n\to K^{+(*)}+e^-]&=&\left|W^{N\to M}_{\hat{\mathcal{Q}}^e_5}C[\hat{\mathcal{Q}}^e_5]\right|^2\nonumber\\&&
+\left|W^{N\to M}_{\hat{\mathcal{Q}}^e_6}C[\hat{\mathcal{Q}}^e_6]\right|^2\,,\nonumber
\end{eqnarray}
\begin{eqnarray}
\tilde{\Gamma}[p\to K^{+(*)}+\nu^{(c)}]&=&\left|W^{N\to M}_{\hat{\mathcal{O}}_1^{\nu}}C[\hat{\mathcal{O}}_1^{\nu}]\right.\nonumber\\
\left.+W^{N\to M}_{\hat{\mathcal{O}}_1'^{\nu}}C[\hat{\mathcal{O}}_1'^{\nu}]\right.
&+&\left.W^{N\to M}_{\hat{\mathcal{R}}^{\nu}}C[\hat{\mathcal{R}}^{\nu}]\right|^2\nonumber\\
+\left|W^{N\to M}_{\hat{\mathcal{Q}}_1^{\nu}}C[\hat{\mathcal{Q}}_1^{\nu}]\right.
&+&\left.W^{N\to M}_{\hat{\mathcal{Q}}'^{\nu}_1}C[\hat{\mathcal{Q}}'^{\nu}_1]\right.\nonumber\\
&+&\left.W^{N\to M}_{\hat{\mathcal{Q}}^{\nu}_2}C[\hat{\mathcal{Q}}^{\nu}_2]\right|^2\nonumber\\
&=&\tilde{\Gamma}[n\to K^{0(*)}+\nu^{(c)}]\,,\nonumber
\end{eqnarray}
where the $N\to M$ superscript corresponds to the nucleon ($N$) to meson ($M$) transition and $\tilde{\Gamma}$ is defined differently for a 
pseudoscalar ($\Pi$) and a vector ($V$) meson:
\begin{eqnarray}
\tilde{\Gamma}[N\to\Pi+L^{(c)}]&\equiv&
\frac{32\pi}{M_N\,\eta^2_{\text{QCD}}}
\left(1-\frac{M_{\Pi}^2}{M_N^2}\right)^{-2}\,,  \nonumber \\
&&\Gamma[N\to\Pi+L^{(c)}]
\end{eqnarray}
\begin{eqnarray}
\tilde{\Gamma}[N\to V+L^{(c)}]&\equiv&\frac{64\pi M_V^2}{M_N^3\,\eta^2_{\text{QCD}}}
\left(1-\frac{M_V^2}{M_N^2}\right)^{-2} \\&& \left(1+2\frac{M_V^2}{M_N^2}\right)^{-1}\Gamma[N\to V+L^{(c)}] \,.\nonumber
\end{eqnarray} 
Here, $\eta_{\text{QCD}}$ represents the QCD running factor of Eq.~(\ref{eqn:running}). 


\section{Bounds on \boldmath$R$-parity-violating parameters}
\label{sec:bounds}

\subsection{Approximations}
The ingredients described in the previous section can be included within a code and would allow to perform a test 
that takes the mixing effects into account to their full extent. Nevertheless, for simplicity, we perform several 
approximations below on the mixings in the SUSY sector and RpV violation, which allow to derive 
analytical bounds on the RpV couplings:
\begin{itemize}
\item we neglect secondary electroweakino-lepton mixing, only allowing for the leading higgsino-lepton mixing generated by the bilinear RpV parameters;
\item we linearize the sfermion-electroweakino/lepton-quark couplings in terms of lepton-violating RpV parameters;
\item we neglect Yukawa couplings of the first generation;
\item we assume that the squark sector is aligned with the Yukawa struture of the quarks, so that the squark mass-matrices do not involve new sources of flavor violation;
\item we neglect left-right squark mixing except for the third generation. 
\end{itemize}
Exploiting these hypotheses considerably simplifies the expression of the Wilson coefficients for the operators of Eq.~(\ref{eq:lowop}):
\begin{eqnarray}\label{eq:Wilsonlep}
C[\mathcal{O}^{e_l}_1]&=&-\lambda''_{1g1}V^{\text{CKM}\,*}_{1f}\left[Y_d^f\delta_{fg}\frac{\mu^*_l}{\mu^*}-\lambda'^*_{lfg}\right]\frac{|X_{kR}^{D_g}|^2}{m^2_{\tilde D_k}} 
\nonumber \\
C[\hat{\mathcal{O}}^{e_l}_1]&=&-\lambda''_{1g2}V^{\text{CKM}\,*}_{1f}\left[Y_d^f\delta_{fg}\frac{\mu^*_l}{\mu^*}-\lambda'^*_{lfg}\right]\frac{|X_{kR}^{D_g}|^2}{m^2_{\tilde D_k}} \nonumber\\
C[\mathcal{O}^{\nu_l}_1]&=&-\lambda''_{1g1}\lambda'^*_{l1g}\frac{|X_{kR}^{D_g}|^2}{m^2_{\tilde D_k}}  \nonumber \\
C[\hat{\mathcal{O}}^{\nu_l}_1]&=&-\lambda''_{1g2}\lambda'^*_{l1g}\frac{|X_{kR}^{D_g}|^2}{m^2_{\tilde D_k}} \nonumber\\
C[\hat{\mathcal{O}}'^{\nu_l}_1]&=&\lambda''_{1g1}\left[Y_d^2\delta_{g2}\frac{\mu^*_l}{\mu^*}-\lambda'^*_{l2g}\right]\frac{|X_{kR}^{D_g}|^2}{m^2_{\tilde D_k}} \\
C[\mathcal{Q}_1^{\nu_l}]&=&-\lambda''_{131}\lambda'^*_{l31}\frac{X_{kR}^{D_3}X_{kL}^{D_3\,*}}{m^2_{\tilde D_k}} 
\nonumber \\
C[\hat{\mathcal{Q}}_1^{\nu_l}]&=&-\lambda''_{132}\lambda'^*_{l31}\frac{X_{kR}^{D_3}X_{kL}^{D_3\,*}}{m^2_{\tilde D_k}} \nonumber \\
C[\hat{\mathcal{Q}}'^{\nu_l}_1]&=&-\lambda''_{131}\lambda'^*_{l32}\frac{X_{kR}^{D_3}X_{kL}^{D_3\,*}}{m^2_{\tilde D_k}} 
\nonumber \\
C[\hat{\mathcal{Q}}_6^{e_l}]&=&\left(\lambda''_{321}-\lambda''_{312}\right)\lambda'^*_{lf1}V^{\text{CKM}}_{3f}\frac{X_{kR}^{U_3}X_{kL}^{U_3\,*}}{m^2_{\tilde U_k}} \nonumber \\
C[\mathcal{O}_5^{e_l}]&=&C[\hat{\mathcal{O}}_5^{e_l}]=
C[\hat{\mathcal{Q}}_2^{\nu_l}]=C[\hat{\mathcal{Q}}_5^{e_l}] 
=C[\hat{\mathcal{R}}^{\nu_l}]=0 \nonumber
\end{eqnarray}
where $l$ is the lepton-flavor index while summation over repeated indices is implicit. The $X_{kR}^{D_g}$ \textit{etc.}
denote the sfermion mixing factors given in Appendix~\ref{ap:mix}. This list can be
further reduced by neglecting sbottom mixing, CKM mixing or the strange-quark mass. We note however that
all the coefficients of Eq.~(\ref{eq:Wilsonlep}) are linearly independent, \textit{i.e.}\ that one can a priori
find directions in parameter space where only one of these coefficients (and any of the non-trivial
ones) is non-zero.

In Eq.~(\ref{eq:Wilsonlep}), we have considered only the four-fermion operators mediating proton decay
that explicitly include a lepton field. As argued before
\cite{Carlson:1995ji,Hoang:1997kf,Bhattacharyya:1998bx}, it is meaningful to consider operators with an
electroweakino field replacing the lepton, either because the electroweakino is light and \textit{e.g.}~long-lived,
or because a heavy electroweakino mediates higher-dimensional operators for proton decay. In the latter
case, an electroweakino-fermion-sfermion coupling with further RpV sfermion-fermion-fermion interaction
would indeed produce dimension $9$ (or higher) operators\footnote{The loop-diagrams of
  Ref.~\cite{Bhattacharyya:1998bx} would be inconsistent in the context of the tree-level matching that we
  perform here and obviously depend on the renormalization conditions defining the RpV
  couplings.}. Here, we specialize in a subsequent coupling of LLE-type, since the hadronic matrix
elements studied in the previous section would not apply if additional quarks were involved\footnote{On
  the other hand, hadronic matrix elements for $B$-violating operators of dimension $9$ would yield
  further limits on RpV parameters.}. Factorizing out this step, we provide the Wilson coefficients for
the operators of Eq.~(\ref{eq:lowop}) where an electroweakino replaces the lepton. The subscript $b$,
$w$, $h_{(\pm)}$ respectively denote bino, wino, higgsino states (with two states of mass $\pm\mu$ in
the neutral case). Electroweakino mixing can be included in the picture by combining the various Wilson
coefficients.
\begin{eqnarray}\label{eq:Wilsongauhig}
C[\mathcal{O}^{w^{-}}_1]&=&-g_2\lambda''_{131}V^{\text{CKM}\,*}_{13}\frac{X_{kR}^{D_3}X_{kL}^{D_3\,*}}{m^2_{\tilde D_k}}\nonumber\\
  C[\hat{\mathcal{O}}^{w^{-}}_1]&=&-g_2\lambda''_{132}V^{\text{CKM}\,*}_{13}\frac{X_{kR}^{D_3}X_{kL}^{D_3\,*}}{m^2_{\tilde D_k}} \nonumber\\
C[\mathcal{O}^{h^{-}}_1]&=&Y_d^gV^{\text{CKM}\,*}_{1g}\lambda''_{1g1}\frac{|X_{kR}^{D_g}|^2}{m^2_{\tilde D_k}}\nonumber\\
  C[\hat{\mathcal{O}}^{h^{-}}_1]&=&Y_d^gV^{\text{CKM}\,*}_{1g}\lambda''_{1g2}\frac{|X_{kR}^{D_g}|^2}{m^2_{\tilde D_k}} \nonumber\\
   C[\hat{\mathcal{O}}'^{h_-}_1]&=&-\frac{Y_d^2}{\sqrt{2}}\lambda''_{121}\frac{|X_{kR}^{D_2}|^2}{m^2_{\tilde D_k}}=-
   C[\hat{\mathcal{O}}'^{h_+}_1]  \nonumber \\
   C[\mathcal{Q}_1^b]&=&-\frac{\sqrt{2}}{3}g_1\lambda''_{111}\frac{|X_{kR}^{D_1}|^2}{m^2_{\tilde D_k}}=0 \nonumber \\
  C[\hat{\mathcal{Q}}^b_1]&=&-\frac{\sqrt{2}}{3}g_1\lambda''_{112}\frac{|X_{kR}^{D_1}|^2}{m^2_{\tilde D_k}}  \\
C[\hat{\mathcal{Q}}'^b_1]&=&-\frac{\sqrt{2}}{3}g_1\lambda''_{121}\frac{|X_{kR}^{D_2}|^2}{m^2_{\tilde D_k}}\nonumber \\
C[\hat{\mathcal{Q}}_2^{b}]&=&\frac{\sqrt{2}}{3}g_1\left(\lambda''_{121}-\lambda''_{112}\right)\frac{|X_{kR}^{U_1}|^2}{m^2_{\tilde U_k}} 
\nonumber \\
C[\hat{\mathcal{Q}}_5^{h^+}]&=&Y_u^gV^{\text{CKM}\,*}_{g1}\left(\lambda''_{g21}-\lambda''_{g12}\right)\frac{|X_{kR}^{U_g}|^2}
{m^2_{\tilde U_k}} \nonumber\\
   C[\mathcal{Q}_1^{w,h_{\pm}}]&=&C[\hat{\mathcal{Q}}_1^{w,h_{\pm}}]=C[\hat{\mathcal{Q}}'^{w,h_{\pm}}_1]=0\nonumber\\
C[\mathcal{O}^{b,w,h_{\pm}}_1]&=&C[\hat{\mathcal{O}}^{b,w,h_{\pm}}_1]=C[\hat{\mathcal{O}}'^{b,w}_1]=0 \nonumber\\
   C[\mathcal{O}_5^{w^-,h^-}]&=&C[\hat{\mathcal{O}}^{w^-,h^-}_5]=0 \nonumber\\
  C[\hat{\mathcal{Q}}_2^{w,h_{\pm}}]&=&C[\hat{\mathcal{Q}}_5^{w^+}]=C[\hat{\mathcal{Q}}_6^{w^+,h^+}]=C[\hat{\mathcal{R}}^{b,w,h}] =0 \nonumber
\end{eqnarray}
Here $g_1=e/\sin\theta_W$ is the $U(1)_Y$ hypercharge gauge coupling and $g_2$ is the $SU(2)_L$ gauge coupling.
Unsurprisingly, the contributions are suppressed (\textit{i.e.}\ require sfermion mixing or vanish) when the
quantum numbers of the bino (SU(2)-singlet), wino (triplet) or higgsinos (doublets) lead to an operator
violating the SM gauge group. However, we note that the operators of type $Q_{1,2}$ with a bino are
SM-conserving, hence fully legitimate dimension $6$ objects. All these coefficients are only
$B$-violating, so that they do not involve $\lambda'$ or $\mu_l$ couplings, in contrast to the
coefficients of Eq.~(\ref{eq:Wilsonlep}). Yet, unless the lightest neutralino is very light, the
`decays' of the electroweakino line require additional RpV effects to contribute to nucleon decays.


\subsection{Nucleon Decays into a Meson and a Lepton}
The experimental results for nucleon decay modes into a meson and an (anti)lepton are collected in \cite{Tanabashi:2018oca}. The channels with pseudoscalar mesons in the final state tend to be more constrained than those with vector mesons. Thus, in consideration of our conservative estimates of the hadronic matrix elements for the decays into vector meson, these latter processes hardly have a chance to compete in the current situation\footnote{However, we note that this would not be systematically the case if we also employed the hadronic matrix elements derived with the bag models for the decays into pseudoscalar mesons, since the branching ratios associated with the vector mesons are then larger. A more precise determination of the hadronic matrix elements for the nucleon to vector meson transition could thus increase the relevance of these channels.}. Therefore, the limits on the Wilson coefficients of Eq.~(\ref{eq:Wilsonlep}) principally derive from the decays into pseudoscalar mesons:
\begin{itemize}
\item from $p\to\pi^0e^+$ \cite{Miura:2016krn}: \\[2mm] \hspace{0.5cm} $\eta_{\text{QCD}}|C[\mathcal{O}_1^e]|<9.2\cdot10^{-32}~\text{GeV}^{-2}$;
\item from $p\to\pi^0\mu^+$ \cite{Miura:2016krn}: \\[2mm] \hspace{0.5cm} $\eta_{\text{QCD}}|C[\mathcal{O}_1^{\mu}]|<1.5\cdot10^{-31}~\text{GeV}^{-2}$;
\item from $n\to\pi^0\nu$ \cite{Abe:2013lua}: \\[2mm] \hspace{0.5cm} $\eta_{\text{QCD}}\left[{\displaystyle\sum_{\nu}}|C[\mathcal{O}_1^{\nu}]|^2+1.05\sum_{\nu}|C[\mathcal{Q}_1^{\nu}]|^2\right]^{1/2}\\[1mm]<3.5\cdot10^{-31}~\text{GeV}^{-2}$;
\item from $p\to K^0e^+$ \cite{Kobayashi:2005pe}: \\[2mm] \hspace{0.5cm} $\eta_{\text{QCD}}|C[\hat{\mathcal{O}}_1^e]|<6.4\cdot10^{-31}~\text{GeV}^{-2}$;
\item from $p\to K^0\mu^+$ \cite{Regis:2012sn}: \\[2mm] \hspace{0.5cm} $\eta_{\text{QCD}}|C[\hat{\mathcal{O}}_1^{\mu}]|<5.3\cdot10^{-31}~\text{GeV}^{-2}$;
\item from $n\to K^+e^-$ \cite{Berger:1991fa}: \\[2mm] \hspace{0.5cm} $\eta_{\text{QCD}}|C[\hat{\mathcal{Q}}_6^e]|<6.4\cdot10^{-30}~\text{GeV}^{-2}$;
\item from $n\to K^+\mu^-$ \cite{Berger:1991fa}: \\[2mm] \hspace{0.5cm} $\eta_{\text{QCD}}|C[\hat{\mathcal{Q}}_6^{\mu}]|<4.5\cdot10^{-30}~\text{GeV}^{-2}$;
\item from $p\to K^+\nu$ \cite{Abe:2014mwa}: \\[1mm] \hspace{0.5cm} \newline
$\eta_{\text{QCD}}\!\left[\sum_{\nu}|C[\hat{\mathcal{O}}'^{\nu}_1]+0.366\,C[\hat{\mathcal{O}}_1^{\nu}]|^2+1.05\cdot
\right.$\\ $\left.{\displaystyle\sum_{\nu}}|C[\hat{\mathcal{Q}}'^{\nu}_1]+0.295\,C[\hat{\mathcal{Q}}^{\nu}_1]|^2\right]^{1/2}\!\!\!<2.0\cdot10^{-31}\,\text{GeV}^{-2}\!\!.$\newline
\end{itemize}

Assuming a universal squark mass $m_{\tilde{Q}}$ and defining the mixing parameter $\Delta^{\tilde{F}_g}_{LR}
\approx-\left[{\cal M}^2_{\tilde{F}_g}\right]_{LR}/m^2_{\tilde Q}$ (with $\left[{\cal M}^2_{\tilde{F}_g}\right]_{LR}$ 
the left-right squared-mass-matrix element, as defined in App.~\ref{ap:mix}) for $\tilde{F}_g=\tilde{U}_g,\tilde{D}_
g$ ($g=3$), we arrive at the following limits on combinations of RpV parameters (with implicit sum on repeated indices):
\begin{widetext}
  \begin{eqnarray}\label{eq:RpVMLbounds}
\left|\lambda''_{1g1}V^{\text{CKM}\,*}_{1f}\left[Y_d^f\delta_{fg}\frac{\mu_1^*}{\mu^*}-\lambda'^*_{1fg}\right]\right|&<&2.9\cdot10^{-26}\left(\frac{m_{\tilde{Q}}}{1~\text{TeV}}\right)^2\frac{3.15}{\eta_{\text{QCD}}} \nonumber\\
\left|\lambda''_{1g1}V^{\text{CKM}\,*}_{1f}\left[Y_d^f\delta_{fg}\frac{\mu_2^*}{\mu^*}-\lambda'^*_{2fg}\right]\right|&<&4.7\cdot10^{-26}\left(\frac{m_{\tilde{Q}}}{1~\text{TeV}}\right)^2\frac{3.15}{\eta_{\text{QCD}}} \nonumber\\
\left[\left|\lambda''_{1g1}\lambda'^*_{l1g}\right|^2+1.05\left|\lambda''_{131}\lambda'^*_{l31}\Delta^{\tilde{D}_3}_{LR}\right|^2\right]^{1/2}&<&1.1\cdot10^{-25}\left(\frac{m_{\tilde{Q}}}{1~\text{TeV}}\right)^2\frac{3.15}{\eta_{\text{QCD}}} \\
\left|\lambda''_{1g2}V^{\text{CKM}\,*}_{1f}\left[Y_d^f\delta_{fg}\frac{\mu_1^*}{\mu^*}-\lambda'^*_{1fg}\right]\right|&<&2.0\cdot10^{-25}\left(\frac{m_{\tilde{Q}}}{1~\text{TeV}}\right)^2\frac{3.15}{\eta_{\text{QCD}}} \nonumber\\
\left|\lambda''_{1g2}V^{\text{CKM}\,*}_{1f}\left[Y_d^f\delta_{fg}\frac{\mu_2^*}{\mu^*}-\lambda'^*_{2fg}\right]\right|&<&1.7\cdot10^{-25}\left(\frac{m_{\tilde{Q}}}{1~\text{TeV}}\right)^2\frac{3.15}{\eta_{\text{QCD}}} \nonumber \\
\left|(\lambda''_{321}-\lambda''_{312})V^{\text{CKM}\,*}_{3f}\lambda'^*_{1f1}\Delta^{\tilde{U}_3}_{LR}\right|&<&2.0\cdot10^{-24}\left(\frac{m_{\tilde{Q}}}{1~\text{TeV}}\right)^2\frac{3.15}{\eta_{\text{QCD}}} \nonumber
\end{eqnarray}
\end{widetext}
\begin{widetext}
  \begin{eqnarray}
\left|(\lambda''_{321}-\lambda''_{312})V^{\text{CKM}\,*}_{3f}\lambda'^*_{2f1}\Delta^{\tilde{U}_3}_{LR}\right|&<&1.4\cdot10^{-24}\left(\frac{m_{\tilde{Q}}}{1~\text{TeV}}\right)^2\frac{3.15}{\eta_{\text{QCD}}} \nonumber \\
\left[\left|\lambda''_{1g1}\left[Y_d^2\delta_{g2}\frac{\mu_2^*}{\mu^*}-\lambda'^*_{l2g}\right]-0.366\lambda''_{1g2}\lambda'^*_{l1g}\right|^2\right.+1.05|\Delta_{LR}^{\tilde{D}_3}|^2
\left|\lambda''_{131}\lambda'^*_{l32}\right.&+&\left.\left.0.295\lambda''_{132}\lambda'^*_{l31}\right|^2 \right]^{1/2}\nonumber\\
\null\hspace{10cm}&<&6.4\cdot10^{-26}\left(\frac{m_{\tilde{Q}}}{1~\text{TeV}}\right)^2\frac{3.15}{\eta_{\text{QCD}}} \nonumber
\end{eqnarray}
\end{widetext}
These constraints update earlier bounds (\textit{e.g.}~\cite{Barbier:2004ez}) with somewhat stronger limits, due in particular
to the relatively recent results from the Super-Kamiokande experiment. However, the theoretical uncertainties have not been 
taken into account in the derivation above. In addition to the uncertainties associated to the hadronic matrix elements (of order 
$10\%$ according to \cite{Aoki:2017puj}), one should add those uncertainties originating  with the Wilson 
coefficients. While the leading QCD logarithms should be properly included in $\eta_{\text{QCD}}$, finite and NLO QCD 
corrections are not considered and could easily amount to a $\sim30\%$ modification. Electroweak logarithms 
have also been neglected, as well as electroweak flavor-changing effects. We believe that the latter can only be consistently 
included by performing a matching of one-loop order and choosing a particular renormalization scheme (so that the RpV 
parameters are set to a specific definition). Finally, we should stress that the approximations on squark and electroweakino-lepton 
mixing over-simplify the system so that the bounds of Eq.~(\ref{eq:RpVMLbounds}) cannot replace a full numerical test. 
In particular, the relevance of neglecting flavor-changing effects in the sfermion sector while allowing flavor-violation in the RpV couplings can be questioned.
Therefore, these limits should be seen at a purely qualitative level.

\subsection{Nucleon Decays into a Meson and a Long-lived Bino}
In this section, we assume the existence of a light bino with mass comparable or below that of the muon and 
long-lived so that it would appear as an invisible particle in nucleon decays. Then, the dimension $6$ operators 
of Eq.~(\ref{eq:Wilsongauhig}) are constrained by the limits on $p\to K^+\nu$.\footnote{In principle, the kinematics
of the decay to a light but massive neutralino are different than to a neutrino. The detailed consideration is beyond the
present work.} In principle, the decay channels into bino should be 
added to those into (anti)neutrino presented in the previous subsection. However, for simplicity, we neglect the purely 
leptonic contribution to invisible decays below, \textit{e.g.}~because $L$-violating parameters vanish.
\begin{itemize}
\item from $p\to K^+\nu$ \cite{Abe:2014mwa}
\end{itemize}

\vspace{-0.7cm}

\begin{eqnarray}
\eta_{\text{QCD}}|C[\hat{\mathcal{Q}}'^b_1]+0.295\,C[\hat{\mathcal{Q}}_1^b]&-&0.705\,C[\hat{\mathcal{Q}}_2^b]|\nonumber \\
&<&2.0\cdot10^{-31}~\text{GeV}^{-2},\nonumber
\end{eqnarray}
from which we deduce (using $\lambda''_{112}=-\lambda''_{121}$):
\begin{align}
& |\lambda''_{121}|<3.9\cdot10^{-25}\left(\frac{m_{\tilde{Q}}}{1~\text{TeV}}\right)^2\frac{3.15}{\eta_{\text{QCD}}}\frac{0.350}{g_1}
\end{align}

\subsection{Nucleon Decays into a Meson and three Leptons}
The gauginos are now assumed to be heavy, they are thus intermediate and off-shell in the decay and can themselves decay to three
leptons. The Wilson coefficients of Eq.~(\ref{eq:Wilsongauhig}) can be combined with gaugino and RpV couplings of LLE type in order to form the following dimension $9$ (or
higher from the $SU(2)_L$-conserving perspective):
 \begin{eqnarray}
  \label{eq:opdim9}
\mathcal{S}_1&=&\varepsilon_{\alpha\beta\gamma}[(\overline{d^c})^{\alpha}P_Ru^{\beta}][(\overline{u^c})^{\gamma}P_Le_l][\overline{e}_{n}P_Le_m] \,,
\nonumber \\
\hat{\mathcal{S}}_1&=&\varepsilon_{\alpha\beta\gamma}[(\overline{s^c})^{\alpha}P_Ru^{\beta}][(\overline{u^c})^{\gamma}P_Le_l][\overline{e}_{n}P_Le_m]\,,\nonumber\\
 \hat{\mathcal{T}}_{11}&=&\varepsilon_{\alpha\beta\gamma}[(\overline{s^c})^{\alpha}P_Ru^{\beta}][(\overline{d^c})^{\gamma}P_R\nu^c_l][\overline{e}_{m}P_Le_n] \,,
\nonumber \\
\hat{\mathcal{T}}'_{11}&=&\varepsilon_{\alpha\beta\gamma}[(\overline{d^c})^{\alpha}P_Ru^{\beta}][(\overline{s^c})^{\gamma}P_R\nu^c_l][\overline{e}_{m}P_Le_n]\,,\nonumber\\
\hat{\mathcal{T}}_{12}&=&\varepsilon_{\alpha\beta\gamma}[(\overline{s^c})^{\alpha}P_Ru^{\beta}][(\overline{d^c})^{\gamma}P_Re_l][\overline{e}_{m}P_L\nu_n] \,,
\nonumber \\
\hat{\mathcal{T}}'_{12}&=&\varepsilon_{\alpha\beta\gamma}[(\overline{d^c})^{\alpha}P_Ru^{\beta}][(\overline{s^c})^{\gamma}P_Re_l][\overline{e}_{m}P_L\nu_n]\,,\nonumber\\
\hat{\mathcal{T}}_{13}&=&\varepsilon_{\alpha\beta\gamma}[(\overline{s^c})^{\alpha}P_Ru^{\beta}][(\overline{d^c})^{\gamma}P_Re_l][\overline{e}_{m}P_R\nu^c_n] \,,
\nonumber \\
\hat{\mathcal{T}}'_{13}&=&\varepsilon_{\alpha\beta\gamma}[(\overline{d^c})^{\alpha}P_Ru^{\beta}][(\overline{s^c})^{\gamma}P_Re_l][\overline{e}_{m}P_R\nu^c_n]\,,\nonumber\\
\hat{\mathcal{T}}_{14}&=&\varepsilon_{\alpha\beta\gamma}[(\overline{s^c})^{\alpha}P_Ru^{\beta}][(\overline{d^c})^{\gamma}P_Re^c_l][\overline{\nu}_{n}P_Re_m] \,,
 \\
\hat{\mathcal{T}}'_{14}&=&\varepsilon_{\alpha\beta\gamma}[(\overline{d^c})^{\alpha}P_Ru^{\beta}][(\overline{s^c})^{\gamma}P_Re^c_l][\overline{\nu}_{n}P_Re_m]\,,\nonumber\\
\hat{\mathcal{T}}_{15}&=&\varepsilon_{\alpha\beta\gamma}[(\overline{s^c})^{\alpha}P_Ru^{\beta}][(\overline{d^c})^{\gamma}P_Re^c_l][\overline{\nu^c}_{n}P_Le_m] \,,
\nonumber \\
\hat{\mathcal{T}}'_{15}&=&\varepsilon_{\alpha\beta\gamma}[(\overline{d^c})^{\alpha}P_Ru^{\beta}][(\overline{s^c})^{\gamma}P_Re^c_l][\overline{\nu^c}_{n}P_Le_m]\,,\nonumber\\
\hat{\mathcal{T}}_{21}&=&\varepsilon_{\alpha\beta\gamma}[(\overline{s^c})^{\alpha}P_Rd^{\beta}][(\overline{u^c})^{\gamma}P_R\nu^c_l][\overline{e}_{m}P_Le_n] \,,
\nonumber \\
\hat{\mathcal{T}}_{22}&=&\varepsilon_{\alpha\beta\gamma}[(\overline{s^c})^{\alpha}P_Rd^{\beta}][(\overline{u^c})^{\gamma}P_Re_l][\overline{e}_{m}P_L\nu_n]\,,\nonumber\\
\hat{\mathcal{T}}_{23}&=&\varepsilon_{\alpha\beta\gamma}[(\overline{s^c})^{\alpha}P_Rd^{\beta}][(\overline{u^c})^{\gamma}P_Re_l][\overline{e}_{m}P_R\nu^c_n] \,,
\nonumber \\
\hat{\mathcal{T}}_{24}&=&\varepsilon_{\alpha\beta\gamma}[(\overline{s^c})^{\alpha}P_Rd^{\beta}][(\overline{u^c})^{\gamma}P_Re^c_l][\overline{\nu}_{n}P_Re_m] \,,
\nonumber\\
\hat{\mathcal{T}}_{25}&=&\varepsilon_{\alpha\beta\gamma}[(\overline{s^c})^{\alpha}P_Rd^{\beta}][(\overline{u^c})^{\gamma}P_Re^c_l][\overline{\nu^c}_{n}P_Le_m]  \,.
\nonumber
\end{eqnarray}
The associated Wilson coefficients read (where we have neglected the Yukawa couplings of the light
leptons and assumed universal slepton masses):
\begin{eqnarray}
C[\mathcal{S}_1]&=&C[\mathcal{O}_1^{w^-}]\frac{g_2\lambda_{lmn}}{M_2m^2_{\tilde{N}_l}} 
\nonumber \\
C[\hat{\mathcal{S}}_1]&=&C[\hat{\mathcal{O}}_1^{w^-}]\frac{g_2\lambda_{lmn}}{M_2m^2_{\tilde{N}_l}} \\
C[\hat{\mathcal{T}}_{11}]&=&-C[\hat{\mathcal{Q}}_1^{b}]\frac{g_1\lambda^*_{lmn}}{\sqrt{2}M_1m^2_{\tilde{N}_l}} 
\nonumber\\
C[\hat{\mathcal{T}}'_{11}]&=&-C[\hat{\mathcal{Q}}_1'^{b}]\frac{g_1\lambda^*_{lmn}}{\sqrt{2}M_1m^2_{\tilde{N}_l}}
\nonumber \\
C[\hat{\mathcal{T}}_{21}]&=&-C[\hat{\mathcal{Q}}_2^{b}]\frac{g_1\lambda^*_{lmn}}{\sqrt{2}M_1m^2_{\tilde{N}_l}}\nonumber 
\end{eqnarray}

\vspace{-0.9cm}

\begin{eqnarray}
C[\hat{\mathcal{T}}_{12}]&=&C[\hat{\mathcal{Q}}_1^{b}]\frac{\sqrt{2}g_1\lambda_{lnm}\Delta_{LR}^{\tilde{E}_l\,*}}{M_1m^2_{\tilde{E}_l}} 
\nonumber \\
C[\hat{\mathcal{T}}'_{12}]&=&C[\hat{\mathcal{Q}}_1'^{b}]\frac{\sqrt{2}g_1\lambda_{lnm}\Delta_{LR}^{\tilde{E}_l\,*}}{M_1m^2_{\tilde{E}_l}} 
\nonumber 
\end{eqnarray}

\vspace{-0.3cm}

\begin{eqnarray}
C[\hat{\mathcal{T}}_{22}]&=&C[\hat{\mathcal{Q}}_2^{b}]\frac{\sqrt{2}g_1\lambda_{lnm}\Delta_{LR}^{\tilde{E}_l\,*}}{M_1m^2_{\tilde{E}_l}}\nonumber\\
C[\hat{\mathcal{T}}_{13}]&=&C[\hat{\mathcal{Q}}_1^{b}]\frac{\sqrt{2}g_1\lambda^*_{nml}}{M_1m^2_{\tilde{E}_l}} 
\nonumber 
\end{eqnarray}
\begin{eqnarray}
C[\hat{\mathcal{T}}_{23}]&=&C[\hat{\mathcal{Q}}_2^{b}]\frac{\sqrt{2}g_1\lambda^*_{nml}}{M_1m^2_{\tilde{E}_l}}\nonumber\\
C[\hat{\mathcal{T}}'_{13}]&=&C[\hat{\mathcal{Q}}_1'^{b}]\frac{\sqrt{2}g_1\lambda^*_{nml}}{M_1m^2_{\tilde{E}_l}} 
\nonumber \\
C[\hat{\mathcal{T}}_{14}]&=&-C[\hat{\mathcal{Q}}_1^{b}]\frac{g_1\lambda^*_{lnm}}{\sqrt{2}M_1m^2_{\tilde{E}_l}} 
\nonumber \\
C[\hat{\mathcal{T}}'_{14}]&=&-C[\hat{\mathcal{Q}}_1'^{b}]\frac{g_1\lambda^*_{lnm}}{\sqrt{2}M_1m^2_{\tilde{E}_l}} 
\nonumber \\
C[\hat{\mathcal{T}}_{24}]&=&-C[\hat{\mathcal{Q}}_2^{b}]\frac{g_1\lambda^*_{lnm}}{\sqrt{2}M_1m^2_{\tilde{E}_l}}\nonumber\\
C[\hat{\mathcal{T}}_{15}]&=&-C[\hat{\mathcal{Q}}_1^{b}]\frac{g_1\lambda_{nml}\Delta_{LR}^{\tilde{E}_l\,*}}{\sqrt{2}M_1m^2_{\tilde{E}_l}} 
\nonumber \\
C[\hat{\mathcal{T}}'_{15}]&=&-C[\hat{\mathcal{Q}}_1'^{b}]\frac{g_1\lambda_{nml}\Delta_{LR}^{\tilde{E}_l\,*}}{\sqrt{2}M_1m^2_{\tilde{E}_l}} 
\nonumber \\
C[\hat{\mathcal{T}}_{25}]&=&-C[\hat{\mathcal{Q}}_2^{b}]\frac{g_1\lambda_{nml}\Delta_{LR}^{\tilde{E}_l\,*}}{\sqrt{2}M_1m^2_{\tilde{E}_l}}\nonumber 
\end{eqnarray}
Here the $m_{\tilde N_l}$ denote the sneutrino masses.
Applying lifetime limits from inclusive nucleon decay channels with antileptons
\cite{Learned:1979gp,Cherry:1981uq}, we derive the following RPV limits (with $m_{\tilde{L}}^2$ the universal slepton mass):
\begin{widetext}
\begin{itemize}
\item from $p\to\pi^0e^+_le^-_me^+_n$: 
\begin{equation}
|\Delta^{\tilde{D}_3}_{LR}\lambda''_{131}|\left[|\lambda_{211}|^2+|\lambda_{122}|^2\right]^{1/2}<2.3\cdot10^{-9}\left(\frac{m_{\tilde{Q}}}{1~\text{TeV}}\right)^2\left(\frac{m_{\tilde{L}}}{1~\text{TeV}}\right)^2\frac{M_2}{1~\text{TeV}}\frac{3.15}{\eta_{\text{QCD}}}\left(\frac{0.653}{g_2}\right)^2 \,,
\end{equation}
\item from $p\to K^0e^+_le^-_me^+_n$: 
\begin{equation}
|\Delta^{\tilde{D}_3}_{LR}\lambda''_{132}|\left[|\lambda_{211}|^2+|\lambda_{122}|^2\right]^{1/2}<2.4\cdot10^{-8}\left(\frac{m_{\tilde{Q}}}{1~\text{TeV}}\right)^2\left(\frac{m_{\tilde{L}}}{1~\text{TeV}}\right)^2\frac{M_2}{1~\text{TeV}}\frac{3.15}{\eta_{\text{QCD}}}\left(\frac{0.653}{g_2}\right)^2\,,
\end{equation}
\item from $p\to K^+\nu^{(c)}_le^-_ne^+_m$: 
\begin{align}
&|\lambda''_{112}\lambda^*_{lm2}|<1.6\cdot10^{-10}\left(\frac{m_{\tilde{Q}}}{1~\text{TeV}}\right)^2\left(\frac{m_{\tilde{L}}}{1~\text{TeV}}\right)^2\frac{M_1}{1~\text{TeV}}\frac{3.15}{\eta_{\text{QCD}}}\left(\frac{0.350}{g_1}\right)^2\,,\nonumber\\
&|\lambda''_{112}\lambda^*_{lm1}|<7.0\cdot10^{-10}\left(\frac{m_{\tilde{Q}}}{1~\text{TeV}}\right)^2\left(\frac{m_{\tilde{L}}}{1~\text{TeV}}\right)^2\frac{M_1}{1~\text{TeV}}\frac{3.15}{\eta_{\text{QCD}}}\left(\frac{0.350}{g_1}\right)^2\,.
\end{align}
\end{itemize}
\end{widetext}
Here, we employed the four-body final-state phase space  derived in \cite{Asatrian:2012tp} and neglected the lepton masses.

\subsection{Purely leptonic final states}
Experimental constraints on purely leptonic decay channels are also available \cite{McGrew:1999nd}. These could a priori be mediated by the 
strangeness-conserving dimension $9$ operator of Eq.~(\ref{eq:opdim9}), $\mathcal{S}_1$. Additional contributions from the operators of Eq.~(\ref{eq:lowop}) 
and an off-shell photon seem difficult to assess in the context of non-perturbative QCD. We will not consider them. We then need to evaluate 
the nucleon decay constant $\left<0\right|\mathcal{S}\left|N\right>$. $\left<0\right|$ represents the QCD vacuum and $N=p,n$, the nucleon. For this, we 
exploit the LEC of the chiral model \cite{Claudson:1981gh} and write:
\begin{eqnarray}
\left<0\right|\varepsilon_{\alpha\beta\gamma}[(\overline{d^c})^{\alpha}P_Ru^{\beta}][(\overline{u^c})^{\gamma}P_L]\left|p\right>&=&\tilde{\alpha}[P_Lu_p]\,,
\\[2mm]
\left<0\right|\varepsilon_{\alpha\beta\gamma}[(\overline{d^c})^{\alpha}P_Ru^{\beta}][(\overline{d^c})^{\gamma}P_R]\left|n\right>&=&\tilde{\beta}[P_Ru_n]\,,
\end{eqnarray}
with $\tilde{\alpha}$ and $\tilde{\beta}$ the LEC calculated in Ref.~\cite{Aoki:2017puj} (see Eq.~(23) of this reference). These quantities are a 
priori valid in the limit of vanishing energy transfer so that their use in decay processes with energy comparable to the nucleon mass is highly 
unreliable. Ref.~\cite{Aoki:2017puj} quotes a factor $2$-$3$ (on the conservative side) in the case of nucleon decay widths into a meson and an 
(anti)lepton, as compared to the full lattice evaluation of the hadronic matrix elements. Thus, we again expect results of purely qualitative value.

The transitions mediated by the operators of type $\mathcal{S}_1$ lead to the following limits:
\begin{widetext}
\begin{itemize}
\item from $p\to e^+\mu^+\mu^-$ \cite{McGrew:1999nd}:
\begin{equation}
|\Delta_{LR}^{\tilde{D}_3}\lambda''_{131}\lambda_{122}|<5.1\cdot10^{-11}\left(\frac{m_{\tilde{Q}}}{1~\text{TeV}}\right)^2\left(\frac{m_{\tilde{L}}}{1~\text{TeV}}\right)^2\frac{M_2}{1~\text{TeV}}\frac{3.15}{\eta_{\text{QCD}}}\left(\frac{0.653}{g_2}\right)^2\,,
\end{equation}
\item from $p\to e^+\mu^+e^-$ \cite{McGrew:1999nd}:
\begin{equation}
|\Delta_{LR}^{\tilde{D}_3}\lambda''_{131}\lambda_{211}|<4.2\cdot10^{-11}\left(\frac{m_{\tilde{Q}}}{1~\text{TeV}}\right)^2\left(\frac{m_{\tilde{L}}}{1~\text{TeV}}\right)^2\frac{M_2}{1~\text{TeV}}\frac{3.15}{\eta_{\text{QCD}}}\left(\frac{0.653}{g_2}\right)^2\,,
\end{equation}
\end{itemize}
\end{widetext}
which, for these specific directions, are stronger than the constraints obtained with the inclusive decay widths including an antilepton.

\section{Conclusions}
In this paper, we have re-visited the constraints from nucleon decays on RpV parameters, updating the bounds with current lattice 
calculations and experimental limits. We have also paid a more detailed attention to the derivation of these constraints than usually 
presented in the literature. Nucleon decays could take a very diverse pattern in the context of RpV and the current sets of bounds 
are restricted by the limited knowledge of hadronic matrix elements. We have exhumed the bag model for an estimate of the 
transitions involving vector mesons, but the outcome suffers from the comparison with the precise lattice results available for decays 
into pseudoscalar mesons. For this reason --- and the associated performance of experimental searches ---, limits from the nucleon 
transition to pseudoscalar meson and (anti)lepton (or invisible) place the most stringent limits on RpV parameters.

In the RpV MSSM, it is also possible to build $L$-conserving $B$-violating operators involving electroweakinos, 
opening further search modes. Once again, the full exploitation of these channels is limited by the absence of 
reliable evaluations of hadronic matrix elements for \textit{e.g.}~purely leptonic nucleon decays or channels 
with multiple mesons in the final state.

On the high-energy side, we have restricted ourselves to a pure tree-level matching of the Wilson coefficients, as a one-loop matching 
would be technically much more involved. As a consequence, the limits that we have derived in Sec.~\ref{sec:bounds} should be seen 
as largely qualitative. In particular, we renounced flavor-violating loops enlarging the set of RpV couplings that can be constrained, as 
sometimes presented in the literature. More precise limits could naturally be derived in an analysis of one-loop order, but
these should then also depend on the renormalization conditions chosen to fix the counterterms of the RpV parameters, a point that seems to have been overlooked in corresponding proposals.

\appendix

\section{MSSM, RpV and Mixing}\label{ap:mix}

\subsection{Mixing in the Squark Sector} 
\label{ap:mix-squark}
This is a R-parity conserving effect. The sfermion mixing matrices can be written in the $(\tilde{F}_L,\tilde{F}_R^{c\,*})$ basis as:
\begin{widetext}
\begin{equation}\label{eqn:Fmat}
{\cal M}^2_{\tilde{F}}=\begin{bmatrix}
m_{F_L}^2+Y_f^2v_f^2+\frac{1}{2}\left(\frac{{\cal Y}_L^f}{2}g_1^2-I_3^fg_2^2\right)(v_u^2-v_d^2)
&Y_fv_f\left(A_f^*-\mu\frac{v_{f'}}{v_f}\right)\\
Y_fv_f\left(A_f-\mu^*\frac{v_{f'}}{v_f}\right)
&m_{F_R}^2+Y_f^2v_f^2+\frac{{\cal Y}_R^f}{4}g_1^2(v_u^2-v_d^2)
\end{bmatrix}
\end{equation}
\end{widetext}
where $f$ is the fermion corresponding to the sfermion $\tilde{F}$, while $f'$ is its $SU(2)_L$ partner. Then, $Y_f$ is the 
associated Yukawa coupling, ${\cal Y}_{L,R}^f$ the associated hypercharges, $I_3^f$, the isospin. Finally, $v_f$ denotes 
the vev.\ of the Higgs doublet to which the fermion $f$ couples at tree-level. In principle, each matrix element in 
Eq.~(\ref{eqn:Fmat}) should be understood as a $3\times3$ block in flavor space. With $M_{\text{SUSY}}$ above the 
electroweak scale, left-right mixing in the squark sector is only relevant for $Y_f=Y_{t,b}$ (but $A_f$ is still a matrix in 
flavor space, meaning that right-handed squarks of the third generation could still have a relevant mixing with left-handed 
squarks of any generation).

We define the (unitary) mixing matrix $X^{\tilde{F}}$, such that ${\cal M}^2_{\tilde{F}}=X^{\tilde{F}\,\dagger}\text{diag}
[m^2_{\tilde{F}_i}]X^{\tilde{F}}$. Then, the gauge eigenstates are connected to the mass eigenstates through: $\tilde{F}_i
=X^{\tilde{F}\,*}_{iL}\tilde{F}_L+X^{\tilde{F}\,*}_{iR}\tilde{F}_R^{c\,*}$, and reciprocally, $\tilde{F}_L=X^{\tilde{F}}_{iL}\tilde{F}_i$, 
$\tilde{F}_R^c=X^{\tilde{F}\,*}_{iR}\tilde{F}^*_i$. The mass matrix of Eq.~(\ref{eqn:Fmat}) should be diagonalized in a fully 
unprejudiced fashion as to the magnitude of the matrix elements, in general. However, it is instructive to consider the 
expansion in terms of the electroweak vev.'s. Then (neglecting inter-generation mixing), $m^2_{\tilde{F}_1}\approx m_
{F_L}^2$, $m^2_{\tilde{F}_2}\approx m_{F_R}^2$, $X^{\tilde{F}}_{1L},X^{\tilde{F}}_{2R}\approx1$ and $X^
{\tilde{F}}_{1R}\approx-X^{\tilde{F}\,*}_{2L}\approx -[{\cal M}^2_{\tilde{F}}]_{LR}/(M^2_{\tilde{F}_1}-M^2_{\tilde{F}_2})$. The 
left-right mixing is obviously associated with an electroweak vev.\ and is thus liable to generate contributions to dimension 
7 operators.

\subsection{Mixing in the Chargino/Lepton Sector} 
This involves both R-parity conserving (wino-higgsino mixing) and R-parity violating effects (higgsino-lepton mixing) \cite{Allanach:2003eb}. 
We work in the description where the sneutrino fields do not take a vev. Then the mass-terms read $-{\cal L}\ni
(\tilde{w}^+,\tilde{h}_u^+,e_R^{c\,f}){\cal M}_{\tilde{C}}(\tilde{w}^-,\tilde{h}_d^-,e_L^g)^T+h.c.$, with:
\begin{equation}\label{eqn:Cmat}
{\cal M}_{\tilde{C}}=\begin{bmatrix}
M_2 & g_2 v_d & 0\\ g_2 v_u & \mu & \mu_g\\ 0 & 0 & Y_e^f\delta_{fg}v_d
\end{bmatrix}\,.
\end{equation}
$f,g$ correspond to the flavor indices. $\mu_g$ is the RpV bilinear coupling. The mass matrix of Eq.~(\ref{eqn:Cmat}) is diagonalized with a pair of unitary matrices so that: ${\cal M}_{\tilde{C}}=V^T\text{diag}(m_{\chi^{\pm}_i})U$. The mass-eigenstates are then defined as: $\chi_i^+=V^*_{iw}\tilde{w}^++V^*_{ih}\tilde{h}_u^++V^*_{ie_f}e_R^{c\,f}$, $\chi_i^-=U^*_{iw}\tilde{w}^-+U^*_{ih}\tilde{h}_d^-+U^*_{ie_f}e_L^{f}$, and the gauge eigenstates can be expressed in terms of the mass eigenstates through inversion.

In the hierarchical context $|M_2|,|\mu|,\big||M_2|-|\mu|\big|\gg gv,|\mu_g|,Y_e^gv_d$, we can approximate these mixing elements by the following expressions:
\begin{eqnarray}
U_{1w},\,V_{1w},\,U_{2h}&\approx&1\,, \nonumber\\ 
V_{2h},\,U_{3e_l},\,V_{3e_l}&\approx&1 \,, \nonumber\\
U_{1h}&\approx&\frac{g_2(v_dM_2^*+v_u\mu)}{|M_2|^2-|\mu|^2}\,, \nonumber\\
V_{1h}&\approx&\frac{g_2(v_d\mu+v_uM_2^*)}{|M_2|^2-|\mu|^2}\,, \nonumber\\
U_{2w}&\approx&-\frac{g_2(v_dM_2+v_u\mu^*)}{|M_2|^2-|\mu|^2}\,, \\
U_{2e_l}&\approx&\frac{\mu_l}{\mu}\text{;  } \nonumber\\
V_{2w}&\approx&-\frac{g_2(v_d\mu^*+v_uM_2)}{|M_2|^2-|\mu|^2}\,,\nonumber\\
%
U_{3h}&\approx&-\frac{\mu^*_l}{\mu^*}\,, \nonumber \\
U_{1e_l},\,V_{1e_l},\,V_{2e_l}&\approx&0\text{;  }\nonumber \\  
U_{3w},\,V_{3w},\,V_{3h}&\approx&0\,, \nonumber
\end{eqnarray}
where the mass-indices $1$, $2$, $3$ respectively refer to mostly wino, higgsino and lepton states.



\subsection{Mixing in the Neutralino/Neutrino sector} 
This is largely comparable to that in the chargino/lepton sector, however it leads to at least one massive neutrino
\cite{Hall:1983id,Dreiner:2010ye,Dreiner:2011ft}. The mass term is of Majorana type and, in the basis  $(\tilde{b}^0,
\tilde{w}^0,\tilde{h}_d^0,\tilde{h}_u^0,\nu_L^f)$, involves the 7x7 matrix:
\begin{equation}\label{eqn:Nmat}
{\cal M}_{\tilde{N}}=\begin{bmatrix}
M_1 & 0 & -\frac{g_1}{\sqrt{2}} v_d & \frac{g_1}{\sqrt{2}} v_u & 0\\
0 & M_2 & \frac{g_2}{\sqrt{2}} v_d & -\frac{g_2}{\sqrt{2}} v_u & 0\\ 
-\frac{g_1}{\sqrt{2}} v_d &\frac{g_2}{\sqrt{2}} v_d  & 0 & -\mu & 0\\ 
\frac{g_1}{\sqrt{2}} v_u & -\frac{g_2}{\sqrt{2}} v_u & -\mu & 0 & -\mu_g\\ 
0 & 0 & 0 & -\mu_g & 0
\end{bmatrix}.
\end{equation}
This matrix is diagonalized via the 7x7 unitary matrix $N$ according to ${\cal M}_{\tilde{N}}=N^T\text{diag}(m_{\chi^0})N$, 
from which we deduce the mass eigenstates $\chi_i^0=N^*_{ib}\tilde{b}+N^*_{iw}\tilde{w}^0+N^*_{ih_d}\tilde{h}_d^0+N^*
_{ih_u}\tilde{h}_u^0+N^*_{i\nu_f}\nu_L^f$, $i=1,\ldots,7$. Again, in a hierarchical context, the mixing elements can be linearized to:
\begin{eqnarray}
N_{1b},N_{2w},N_{5\nu_l}&\approx&1\,,\qquad   \\
N_{1w},N_{1\nu_l},N_{2b}, N_{2\nu_l}&\approx& 0\,,\nonumber\\
N_{5b},\,N_{5w},\,N_{5h_u} &\approx& 0\nonumber
\end{eqnarray}
\begin{eqnarray}
N_{1h_d}&\approx&-\frac{g_1}{\sqrt{2}}\frac{M_1^*v_d+\mu v_u}{|M_1|^2-|\mu|^2} \nonumber\\
N_{1h_u}&\approx&\frac{g_1}{\sqrt{2}}\frac{M_1^*v_u+\mu v_d}{|M_1|^2-|\mu|^2}\nonumber\\[2mm]
N_{2h_d}&\approx&\frac{g_2}{\sqrt{2}}\frac{M_2^*v_d+\mu v_u}{|M_2|^2-|\mu|^2}\nonumber \\
N_{2h_u}&\approx&-\frac{g_2}{\sqrt{2}}\frac{M_2^*v_u+\mu v_d}{|M_2|^2-|\mu|^2}\,,\nonumber\\
N_{3b}&\approx&\frac{g_1}{2}\frac{(v_d-v_u)(M_1-\mu^*)}{|M_1|^2-|\mu|^2} \nonumber \\
N_{3w}&\approx&\frac{g_2}{2}\frac{(v_u-v_d)(M_2-\mu^*)}{|M_2|^2-|\mu|^2}\nonumber 
\end{eqnarray}
\vspace{-0.3cm}
\begin{eqnarray}
N_{3h_d},N_{3h_u},N_{4h_u},-N_{4h_d}&\approx&\frac{1}{\sqrt{2}}\,,\nonumber\\
N_{3\nu_l},-N_{4\nu_l}&\approx&\frac{\mu_l}{\sqrt{2}\mu}\nonumber\\
N_{4b}&\approx&-\frac{g_1}{2}\frac{(v_d+v_u)(\mu^*+M_1)}{|M_1|^2-|\mu|^2}\nonumber\\ 
N_{4w}&\approx&\frac{g_2}{2}\frac{(v_d+v_u)(\mu^*+M_2)}{|M_2|^2-|\mu|^2}\nonumber \\
N_{5h_d}&\approx&-\frac{\mu_l^*}{\mu^*}\,,
\end{eqnarray}
where the indices $1$, $2$, $3$, $4$ and $5$ correspond to mostly bino, wino, a pair of higgsino and
neutrino states (in fact $5$ covers three leptonic states).


\section{Feynman Rules}\label{ap:FR}

Below, we write the Weyl two-component spinors \cite{Dreiner:2008tw} with lower case letters and the four-component spinors with 
capital letters. The baryon-number-violating couplings involving sups read ($\lambda''_{mnp}\equiv-\lambda''_{mpn}$): 
\begin{eqnarray}
{\cal L}&\ni& \varepsilon_{\alpha\beta\gamma}\lambda''^*_{mnp}(\tilde{U}_R^{c})^{\alpha\,*}_m(\bar{d}_R^c)^{\beta}_n(\bar{d}_R^c)^{\gamma}_p+h.c. 
\label{eq:bnv-sups}\\
&\to& \varepsilon_{\alpha\beta\gamma}\lambda''^*_{rnp}X^{U_r}_{mR}\,\tilde{U}_m^{\alpha}[(\overline{D^c})^{\beta}_nP_R(D)^{\gamma}_p]+h.c.\,,
\end{eqnarray}
and for sdowns
\begin{eqnarray}
{\cal L}&\ni&+\varepsilon_{\alpha\beta\gamma}\lambda''^*_{mnp}(\bar{u}_R^{c})^{\alpha}_m(\bar{d}_R^c)^{\beta}_n(\tilde{D}_R^c)^{\gamma}_p+h.c.
\label{eq:bnv-sdowns}\\
&\to& \varepsilon_{\alpha\beta\gamma}\lambda''^*_{mpr}X^{D_r}_{nR}\,\tilde{D}_n^{\beta}[(\overline{U^c})^{\alpha}_mP_R(D)^{\gamma}_p]+h.c.
\end{eqnarray}
Here $X^{U_r}_{mR},X^{D_r}_{nR}$ denote the squark mixing coefficients, \textit{cf.} Appendix~\ref{ap:mix-squark}. We will use the notations 
$\left(g_R^{Udd}\right)_{mnp}$ and $\left(g_R^{uDd}\right)_{mnp}$ to denote the complete coefficients in the second lines of Eqs.~(\ref{eq:bnv-sups}),
and (\ref{eq:bnv-sdowns}), respectively. We have furthermore in the Lagrangians put the fields in parentheses.
\newline The lepton-number-violating couplings involving sups, downs and charginos read:
\begin{eqnarray}
{\cal L}&\ni& Y_d^f(\tilde{U}_L)^{\alpha*}_f(\bar{\tilde{h}}^+_d)(\bar{d}_R^c)_f^{\alpha}
+\lam'^*_{fgh}(\tilde{U}_L)^{\alpha*}_g(\bar{e}_L)_f(\bar{d}_R^c)_h^{\alpha}\nonumber\\
&+&Y_u^f(\tilde{U}_R^c)_{f}^{\alpha}(\tilde{h}_u^+)(d_L)^{\alpha}_f\nonumber\\
&-&g_2(\tilde{U}_L)_{f}^{\alpha*}(\tilde{w}^+)(d_L)^{\alpha}_f+h.c.\\
&& \hspace{-1cm}\to\tilde{U}^{*\alpha}_m\left\{(\overline{D^c})^{\alpha}_f[g_L^{Ud\chi}P_L+g_R^{Ud\chi}P_R]_{mfq}(\chi^+)_q\right\}+h.c.\nonumber\\
 g_{L\,mfq}^{Ud\chi}&\equiv& V^{\text{CKM}}_{gf}\left[Y_u^gX_{mR}^{U_g\,*}V_{qh}-g_2X_{mL}^{U_g\,*}V_{qw}\right];\\ 
 g_{R\,mfq}^{Ud\chi}&\equiv &
V^{\text{CKM}}_{gr}\left[Y_d^f\delta_{fr}X_{mL}^{U_g\,*}U^*_{qh}+\lambda'^*_{lgf}X_{mL}^{U_r\,*}U^*_{qe_l}\right]\!.
\end{eqnarray}
$g_{1,2}$ are gauge couplings. The lepton-number-violating couplings involving sdowns, ups and charginos read:
\begin{eqnarray}
{\cal L}&\ni& Y_d^f(\tilde{D}_R^c)^{\alpha}_f(\tilde{h}_d^-)(u_L)^{\alpha}_f-g_2(\tilde{D}_L)^{\alpha\,*}(\tilde{w}^-)(u_L)^{\alpha}_f\nonumber \\
&+&\lambda'_{fgk}(\tilde{D}_R^c)^{\alpha}_k(e_L)_{f}(u_L)^{\alpha}_g\nonumber\\
&+&Y_u^f(\tilde{D}_L)^{\alpha\,*}(\bar{h}_u^-)(\bar{u}_R^c)_f^{\alpha}+h.c.\\
&\to& \tilde{D}^{*\,\alpha}_m\left\{(\overline{U^c})^{\alpha}_f[g_L^{Du\chi}P_L+g_R^{Du\chi}P_R]_{mfq}(\chi^-)_q\right\}\nonumber\\
&&+h.c.\\
g_{L\,mfq}^{Du\chi}&=&V^{\text{CKM}\,*}_{fg}\left[Y_d^gX_{mR}^{D_g\,*}U_{qh}-g_2X_{mL}^{D_g\,*}U_{qw}\right.\nonumber \\
&&\left.+\lambda'^*_{lgk}X_{mR}^{D_k\,*}U_{qe_l}\right]\\ 
g_{R\,mfq}^{Du\chi}&=&V^{\text{CKM}\,*}_{fg}Y_u^fX_{mL}^{D_g\,*}V^*_{qh}
\end{eqnarray}
The lepton-number-violating couplings involving sups, ups and neutralinos read:
\begin{widetext}
\begin{align}
{\cal L}\ni& -Y_u^f(\tilde{U}_R^c)_f^{\alpha}(\tilde{h}_u^0)(u_L)_f^{\alpha}-\frac{1}{\sqrt{2}}(\tilde{U}_L)_f^{\alpha\,*}\left[\frac{g_1}{3}(\tilde{b})+g_2(\tilde{w}^0)\right](u_L)_f^{\alpha}
-Y_u^f(\tilde{U}_L)_f^{\alpha\,*}(\bar{\tilde{h}}_u^0)(\bar{u}_R^c)_f^{\alpha}
+\frac{2\sqrt{2}}{3}g_1(\tilde{U}_R^c)_f^{\alpha}(\bar{\tilde{b}})(\bar{u}_R^c)_f^{\alpha}\nonumber \\
&+h.c.\\ 
& \null\hspace{0.1cm}\to \tilde{U}^{*\,\alpha}_m\left\{(\overline{U^c})^{\alpha}_f[g_L^{Uu\chi}P_L+g_R^{Uu\chi}P_R]_{mfq}(\chi^0)_q\right\}+h.c.\\
&g_{L\,mfq}^{Uu\chi}=-Y_u^fX_{mR}^{U_f\,*}N_{qh_u}-\frac{1}{\sqrt{2}}X_{mL}^{U_f\,*}\left[\frac{g_1}{3}N_{qb}+g_2N_{qw}\right]\,,\\
&g_{R\,mfq}^{Uu\chi}=-Y_u^fX_{mL}^{U_f\,*}N^*_{qh_u}+\frac{2\sqrt{2}}{3}g_1X_{mR}^{U_f\,*}N^*_{qb}
\end{align}
The lepton-number-violating couplings involving sdowns, downs and neutralinos read:
\begin{eqnarray}
{\cal L}&\ni& -Y_d^f(\tilde{D}_R^c)^{\alpha}_f(\tilde{h}_d^0)(d_L)_f^{\alpha}-\frac{1}{\sqrt{2}}(\tilde{D}_L)_f^{\alpha\,*}\left[\frac{g_1}{3}(\tilde{b})-g_2(\tilde{w}^0)\right](d_L)_f^{\alpha}-\lambda'_{fgk}(\tilde{D}_R^c)^{\alpha}_k(\nu_L)_f(d_L)^{\alpha}_g -Y_d^f(\tilde{D}_L)_f^{\alpha\,*}(\bar{\tilde{h}}_d^0)(\bar{d}_R^c)_f^{\alpha}
\nonumber\\ 
&&
-\frac{\sqrt{2}}{3}g_1(\tilde{D}_R^c)_f^{\alpha}(\bar{\tilde{b}})(\bar{d}_R^c)_f^{\alpha}-\lambda'^*_{fgk}(\tilde{D}_L)_g^{\alpha\,*}(\bar{\nu}_L)_f(\bar{d}_R^c)_k^{\alpha}+h.c.\nonumber\\
&\to& \tilde{D}^{*\,\alpha}_m\left\{(\overline{D^c})^{\alpha}_f[g_L^{Dd\chi}P_L+g_R^{Dd\chi}P_R]_{mfq}(\chi^0)_q\right\}+h.c.\\
&&g_{L\,mfq}^{Dd\chi}=-Y_d^fX_{mR}^{D_f\,*}N_{qh_d}-\frac{1}{\sqrt{2}}X_{mL}^{D_f\,*}\left[\frac{g_1}{3}N_{qb}-g_2N_{qw}\right]-\lambda'_{gfk}X_{mR}^{D_k\,*}N_{q\nu_g}\,,\nonumber\\ &&g_{R\,mfq}^{Dd\chi}=-Y_d^fX_{mL}^{D_f\,*}N^*_{qh_d}-\frac{\sqrt{2}}{3}g_1X_{mR}^{D_f\,*}N^*_{qb}-\lambda'^*_{gkf}X_{mL}^{D_k\,*}N^*_{q\nu_g}\,.\nonumber
\end{eqnarray}
Omitting the slepton-Higgs mixing, the slepton-lepton/electroweakino couplings read:
\begin{align}
&g^{\tilde{N}\chi^+\chi^-}_{L\,mjk}=Y_e^fX^{\tilde{N}_f}_{mL}V_{je_f}U_{kd}-g_2X^{\tilde{N}_f\,*}_{mL}V_{jw}U_{ke_f}-\lambda_{fpq}X^{\tilde{N}_f}_{mL}V_{je_q}U_{ke_p}=\left(g^{\tilde{N}\chi^+\chi^-}_{R\,mkj}\right)^*\nonumber\\
&g^{\tilde{N}\chi^0\chi^0}_{L\,mjk}=\frac{g_1}{\sqrt{2}}X^{\tilde{N}_f\,*}_{mL}(N_{j\nu_f}N_{kb}+N_{jb}N_{k\nu_f})-\frac{g_2}{\sqrt{2}}X^{\tilde{N}_f\,*}_{mL}(N_{j\nu_f}N_{kw}+N_{jw}N_{k\nu_f})=\left(g^{\tilde{N}\chi^0\chi^0}_{R\,mkj}\right)^*\nonumber\\
&g^{\tilde{E}^*\chi^0\chi^-}_{L\,mjk}=Y_e^fX_{mR}^{\tilde{E}_f\,*}(N_{j\nu_f}U_{kd}-N_{jd}U_{ke_f})+\frac{X_{mL}^{\tilde{E}_f\,*}}{\sqrt{2}}\left[(g_1N_{jb}+g_2N_{jw})U_{ke_f}-g_2N_{j\nu_f}U_{kw}\right]\,,\nonumber\\
&\null\hspace{1.7cm}-\lambda_{fpq}X_{mR}^{\tilde{E}_q\,*}N_{j\nu_f}U_{ke_q}=\left(g^{\tilde{E}\chi^+\chi^0}_{R\,mkj}\right)^*\,,\nonumber\\
&g^{\tilde{E}^*\chi^0\chi^-}_{R\,mjk}=-(Y_e^fX_{mL}^{\tilde{E}_f\,*}N_{jd}^*+\sqrt{2}g_1X_{mR}^{\tilde{E}_f\,*}N_{jb}^*)V^*_{ke_f}-\lambda^*_{fpq}X_{mL}^{\tilde{E}_f\,*}N_{j\nu_p}^*V^*_{ke_q}=\left(g^{\tilde{E}\chi^+\chi^0}_{L\,mkj}\right)^*\,.
\end{align}
\end{widetext}


\section{Static Bag Approach to Nucleon Decays}
In the MIT bag description of hadrons \cite{Chodos:1974je,Chodos:1974pn}, valence quarks are
relativistic fermions trapped in a spherical potential well of radius $R$ (we restrict ourselves to the
flat infinite potential: $V(|\vec{x}|<R)=0$ and $V(|\vec{x}|>R)=\infty$), the boundary of which is
stabilized by a pressure term. The associated fields can then be decomposed in modes:
\begin{eqnarray}
q(x)&=&\sum_{m,s}\left[a^q_{m,s}\,U_{m,s}(\vec{x})e^{-i\omega_m
    t}+b_{m,s}^{q\,\dagger}V_{m,s}(\vec{x})e^{i\omega_m t}\right]\nonumber
\end{eqnarray}
 with
$s=\pm\frac{1}{2}={\mathbin\uparrow\hspace{-.3em}\downarrow}$ the spin and $m$ indexing the solutions 
of the boundary conditions. $a^q_{m,s}$ and $b_{m,s}^{q\,\dagger}$ are creation and 
destruction operators of
(anti)quarks. $\omega_m\equiv E_m/R$ with $E_m$ denoting the energy of the mode. We will restrict
ourselves to the mode of lowest energy,\!\footnote{In the case of the strange quark, we include a quark
  mass --- see Ref.~\cite{Donoghue:1975yg} -- of $0.1$~GeV, which however has negligible impact as compared
  to the massless case.} which can be described by the four-spinors in Dirac representation:
\begin{eqnarray}
U_{0,s}(x)&=&i\sqrt{\frac{\omega_0^3}{8\pi R^3(\omega_0-1)\sin^2\omega_0}}\nonumber \\
&&\begin{pmatrix}
j_0\left(\omega_0\frac{|\vec{x}|}{R}\right)\chi_s\\ i\frac{\vec{x}\cdot\vec{\sigma}}{|\vec{x}|}j_1\left(\omega_0\frac{|\vec{x}|}{R}\right)\chi_s
\end{pmatrix}
\end{eqnarray} 
\begin{eqnarray}
V_{0,s}(x)&=&CU_{0,s}^*\nonumber \\
&=&-i\sqrt{\frac{\omega_0^3}{8\pi R^3(\omega_0-1)\sin^2\omega_0}}\nonumber \\
&&\begin{pmatrix}
-i\frac{\vec{x}\cdot\vec{\sigma}}{|\vec{x}|}j_1\left(\omega_0\frac{|\vec{x}|}{R}\right)\chi'_s\\ j_0\left(\omega_0\frac{|\vec{x}|}{R}\right)\chi'_s
\end{pmatrix}
\end{eqnarray} 
with $j_0(x)=\frac{\sin x}{x}$ and $j_1(x)=\frac{\sin x}{x^2}-\frac{\cos x}{x}$ the first two spherical Bessel functions and $\omega_0\approx2.04$ the first root of the equation $j_0(\omega)=j_1(\omega)$. $\chi_{\uparrow}=\begin{pmatrix}1\\0\end{pmatrix}=-\chi'_{\downarrow}$, $\chi_{\downarrow}=\begin{pmatrix}0\\1\end{pmatrix}=\chi'_{\uparrow}$. 

The hadronic bag states can be constructed with creation operators of the valence quarks $a^{q\,\dagger}_{\alpha s}$ and antiquarks $b^{q\,\dagger}_{\alpha s}$ ($\alpha$ is the color index) acting on the vacuum and satisfying the usual anticommutation relations. For example:
\begin{align}
& \text{proton:}  \nonumber \\ 
& \left|p_{\uparrow}\right>=\frac{\varepsilon_{\alpha\beta\gamma}}{3\sqrt{2}}\,a^{u\,\dagger}_{\alpha\uparrow}(a^{u\,\dagger}_{\beta\uparrow}a^{d\,\dagger}_{\gamma\downarrow}-a^{u\,\dagger}_{\beta\downarrow}a^{d\,\dagger}_{\gamma\uparrow})\left|0\right>\nonumber
\end{align}
\begin{align}
& \text{neutron:} \nonumber \\ 
& \left|n_{\uparrow}\right>=-\frac{\varepsilon_{\alpha\beta\gamma}}{3\sqrt{2}}\,a^{d\,\dagger}_{\alpha\uparrow}(a^{d\,\dagger}_{\beta\uparrow}a^{u\,\dagger}_{\gamma\downarrow}-a^{d\,\dagger}_{\beta\downarrow}a^{u\,\dagger}_{\gamma\uparrow})\left|0\right>\\ 
& \text{neutral pion:} \nonumber \\ 
& \left|\pi^0\right>=\frac{1}{2\sqrt{3}}\left(b^{d\,\dagger}_{\alpha\uparrow}a^{d\,\dagger}_{\alpha\downarrow}-b^{u\,\dagger}_{\alpha\uparrow}a^{u\,\dagger}_{\alpha\downarrow}-b^{d\,\dagger}_{\alpha\downarrow}a^{d\,\dagger}_{\alpha\uparrow}+b^{u\,\dagger}_{\alpha\downarrow}a^{u\,\dagger}_{\alpha\uparrow}\right)\left|0\right>\nonumber\\ 
& \text{neutral rho:} \nonumber \\ 
&\begin{cases}
\left|\rho^0_1\right>=\frac{1}{\sqrt{6}}\left(b^{u\,\dagger}_{\alpha\uparrow}a^{u\,\dagger}_{\alpha\uparrow}-b^{d\,\dagger}_{\alpha\uparrow}a^{d\,\dagger}_{\alpha\uparrow}\right)\left|0\right>\\
\left|\rho^0_0\right>=\frac{1}{2\sqrt{3}}\left(b^{u\,\dagger}_{\alpha\uparrow}a^{u\,\dagger}_{\alpha\downarrow}-b^{d\,\dagger}_{\alpha\uparrow}a^{d\,\dagger}_{\alpha\downarrow}+b^{u\,\dagger}_{\alpha\downarrow}a^{u\,\dagger}_{\alpha\uparrow}-b^{d\,\dagger}_{\alpha\downarrow}a^{d\,\dagger}_{\alpha\uparrow}\right)\left|0\right>\\
\left|\rho^0_{-1}\right>=\frac{1}{\sqrt{6}}\left(b^{u\,\dagger}_{\alpha\downarrow}a^{u\,\dagger}_{\alpha\downarrow}-b^{d\,\dagger}_{\alpha\downarrow}a^{d\,\dagger}_{\alpha\downarrow}\right)\left|0\right>
\end{cases}\nonumber
\end{align}
Then, the matrix element of a partonic operator $\Omega$ between hadronic external states $\left<H_f\right|\int{d\vec{x}\,\Omega(\vec{x})}\left|H_i\right>$ at $t=0$ can be evaluated from replacing the quark fields within $\Omega$ by their expression in the bag model, leading to the usual interplay of Wick contractions. Different bags are employed for the various hadrons, the typical radius being $5\,\text{GeV}^{-1}$ for a nucleon and $3.3\,\text{GeV}^{-1}$ for a pion. A Wick contraction between an external creation/annihilation operator and an internal quark field thus exports the bag wave-function of the corresponding hadron under the $\int{d\vec{x}}$. Contractions between operators involving both external hadrons produce spectator quarks, leading to a separate integral representing the overlap between the two bag functions: for instance, $\left<0\right|a^{q}_{\alpha\uparrow}[H_f]\,a^{q\,\dagger}_{\beta\uparrow}[H_i]\left|0\right>=\int{d\vec{y}\,U^{H_f\,\dagger}_{\uparrow}(\vec{y})U^{H_i}_{\uparrow}(\vec{y})}\delta_{\alpha\beta}$. Below, we detail the case of the $p_{\uparrow}\to\rho^0_{1}e^+$ transition mediated by an operator $\Omega_{\Gamma\Gamma'}=\varepsilon_{\alpha\beta\gamma}[(\overline{d^c})^{\alpha}\Gamma u^{\beta}][(\overline{u^c})^{\gamma}\Gamma'e]$, with $\Gamma$, $\Gamma'$ representing generic spinor-algebra matrices.
\begin{widetext}
\begin{align}
&\left<\rho^0_1\right|\Omega_{\Gamma\Gamma'}\left|p_{\uparrow}\right>=\left<0\right|\frac{1}{\sqrt{6}}\left(b^{u}_{\alpha\uparrow}a^{u}_{\alpha\uparrow}-b^{d}_{\alpha\uparrow}a^{d}_{\alpha\uparrow}\right)\int{d\vec{x}\,\varepsilon_{mnl}[(\overline{d^c})^{m}\Gamma u^{n}][(\overline{u^c})^{l}\Gamma'e]}\frac{\varepsilon_{\beta\gamma\delta}}{3\sqrt{2}}\,a^{u\,\dagger}_{\beta\uparrow}(a^{u\,\dagger}_{\gamma\uparrow}a^{d\,\dagger}_{\delta\downarrow}-a^{u\,\dagger}_{\gamma\downarrow}a^{d\,\dagger}_{\delta\uparrow})\left|0\right>\\
&\null\hspace{0.2cm}=-\frac{1}{\sqrt{3}}\int{d\vec{y}\,U^{\rho\,\dagger}_{\uparrow}(\vec{y})U^{p}_{\uparrow}(\vec{y})}\left\{2\int{d\vec{x}\,[\overline{V}^{p}_{\downarrow}(\vec{x})\Gamma U^{p}_{\uparrow}(\vec{x})][\overline{U}^{\rho}_{\uparrow}(\vec{x})\Gamma' e(\vec{x})]}+2\int{d\vec{x}\,[\overline{V}^{p}_{\downarrow}(\vec{x})\Gamma V^{\rho}_{\uparrow}(\vec{x})][\overline{V}^{p}_{\uparrow}(\vec{x})\Gamma' e(\vec{x})]}\right.\nonumber\\
&\null\hspace{4.3cm}-\int{d\vec{x}\,[\overline{V}^{p}_{\uparrow}(\vec{x})\Gamma V^{\rho}_{\uparrow}(\vec{x})][\overline{V}^{p}_{\downarrow}(\vec{x})\Gamma' e(\vec{x})]}-\int{d\vec{x}\,[\overline{V}^{p}_{\uparrow}(\vec{x})\Gamma U^{p}_{\downarrow}(\vec{x})][\overline{U}^{\rho}_{\uparrow}(\vec{x})\Gamma' e(\vec{x})]}\nonumber\\
&\null\hspace{4.3cm}\left.+\int{d\vec{x}\,[\overline{U}^{\rho}_{\uparrow}(\vec{x})\Gamma U^{p}_{\downarrow}(\vec{x})][\overline{V}^{p}_{\uparrow}(\vec{x})\Gamma' e(\vec{x})]}+\int{d\vec{x}\,[\overline{U}^{\rho}_{\uparrow}(\vec{x})\Gamma U^{p}_{\uparrow}(\vec{x})][\overline{V}^{p}_{\downarrow}(\vec{x})\Gamma' e(\vec{x})]}\right\}\nonumber
\end{align}
\end{widetext}

The connection between this calculation in the bag model and the transition amplitude is not completely
trivial and requires resorting to the wavepacket formalism \cite{Donoghue:1979pr,Donoghue:1979ax}. We
apply the conversion factor in Eq.~(12) of Ref.\cite{Donoghue:1979pr}. The outgoing lepton is regarded
as free, so that its position dependence would be a simple $e^{-i\vec{k}\cdot\vec{x}}$, with
$\vec{k}$ the associated momentum. However, in the static approximation, the frequency $|\vec{k}|$ leads
to a very slow variation, hence this factor can be discarded, or kept
\cite{Golowich:1980ne,Wakano:1982sk} \textit{e.g.}\ in an attempt to extend the prediction to the case of light
pions. In this latter case, however, the static cavity description is not really suited and tentative
corrections should be seen as largely heuristic, such as the phenomenological suppression introduced in
\cite{Donoghue:1979pr}. In our analysis, however, the decay channels into pions are already covered by
the lattice description, so that the results of the bag model is only employed in the more suitable
configuration with heavy mesons in the final state. To complete the calculation of the transition
amplitude, we provide the free-lepton spinors in the Dirac representation:
\begin{eqnarray}
u_s^{\ell}(\vec{k})&=&\begin{pmatrix}\sqrt{E_{\vec{k}}+m_{\ell}}\,\chi_s\\\frac{\vec{k}\cdot\vec{\sigma}}{\sqrt{E_{\vec{k}}+m_{\ell}}}\,\chi_s\end{pmatrix}\,, \\
v_s^{\ell}(\vec{k})&=&\begin{pmatrix}\frac{\vec{k}\cdot\vec{\sigma}}{\sqrt{E_{\vec{k}}+m_{\ell}}}\,\chi_s'\\\sqrt{E_{\vec{k}}+m_{\ell}}\,\chi_s'\end{pmatrix}\,, \\
E_{\vec{k}}&\equiv&\sqrt{\vec{k}^2+m^2_{\ell}}\,;\\
w_s^{\nu}(\vec{k})&=&\sqrt{|\vec{k}|}\begin{pmatrix}\chi_s\\-\chi_s\end{pmatrix}\,,
\end{eqnarray}
with the standard normalization convention. Once all the matrix elements have been computed, it is possible to match them onto the form factors of the decay, \textit{e.g.}: 
\begin{eqnarray}
{\cal A}^{\Omega}[p\to\rho^0e^+]&\equiv&\left<\rho^0,e^+\right|C_{\Omega}\Omega\left|p\phantom{\rho^0\hspace{-0.3cm}}\right> \\
&\equiv& W^{p\to\rho^0}_{[\Omega]}\,\overline{v^c_e}(\vec{k})P_{\Omega}\gamma^{\mu}u_{p}(\vec{0})\,\epsilon^{\rho\,*}_{\mu}(-\vec{k})\,. \nonumber
\end{eqnarray}







\section*{Acknowledgements}
We acknowledge support of the DFG grant SFB CRC-110 {\it Symmetries and the Emergence of Structure in QCD}.
N.C. acknowledges support of the Alexander von Humboldt Foundation and is grateful for the kind hospitality of
the Bethe Center for Theoretical Physics at Bonn University.

\end{document}